\documentclass[oldversion]{aa}

\usepackage{graphicx}
\usepackage{txfonts}
\usepackage{hyperref}
\usepackage{tabularx}

\newcommand{\be}{\begin{equation}}
\newcommand{\ee}{\end{equation}}

 {\everymath{\displaystyle\everymath{}}\array}%
 {\endarray}

\begin{document}

\title{Alleviating the transit timing variation bias in transit surveys\thanks{Table 3 is only available in electronic form at the CDS via anonymous ftp to cdsarc.u-strasbg.fr (130.79.128.5) or via \url{http://cdsweb.u-strasbg.fr/cgi-bin/qcat?J/A+A/ }}
}
\subtitle{I. RIVERS: Method and detection of a pair of resonant super-Earths around Kepler-1705}
\titlerunning{I. RIVERS: Method and detection of a pair of resonant super-Earth around Kepler-1705}

\author{
A. Leleu$^{1,2}$, G. Chatel$^3$, S. Udry$^{1}$, Y. Alibert$^{2}$, J.-B. Delisle$^{1}$ and R. Mardling$^4$}
\authorrunning{A. Leleu et al}

\institute{
$^1$ Observatoire de Gen\`eve, Universit\'e de Gen\`eve, Chemin Pegasi, 51, 1290 Versoix, Switzerland.\\
$^2$ Physikalisches Institut, Universit\"at Bern, Gesellschaftsstr.\ 6, 3012 Bern, Switzerland.\\
$^3$ Disaitek, www.disaitek.ai.\\
$^4$ School of Physics and Astronomy, Monash University, Victoria 3800, Australia.
}

\abstract
{
Transit timing variations (TTVs) can provide useful information for systems observed by transit, as they allow us to put constraints on the masses and eccentricities of the observed planets, or even to constrain the existence of non-transiting companions. However, TTVs can also act as a detection bias that can prevent the detection of small planets in transit surveys that would otherwise be detected by standard algorithms such as the Boxed Least Square algorithm (BLS) if their orbit was not perturbed. This bias is especially present for surveys with a long baseline, such as Kepler, some of the TESS sectors, and the upcoming PLATO mission. 
 Here we introduce a detection method that is robust to large TTVs, and illustrate its use by recovering and confirming a pair of resonant super-Earths with ten-hour TTVs around Kepler-1705 (prev. KOI-4772). The method is based on a neural network trained to recover the tracks of low-signal-to-noise-ratio(S/N) perturbed planets in river diagrams. %
We recover the transit parameters of these candidates by fitting the light curve. The individual transit S/N of Kepler-1705b and c are about three times lower than all the previously known planets with TTVs of 3 hours or more, pushing the boundaries in the recovery of these small, dynamically active planets.
Recovering this type of object is essential for obtaining a complete picture of the observed planetary systems, and solving for a bias not often taken into account in statistical studies of exoplanet populations. In addition, TTVs are a means of obtaining mass estimates which can be essential for studying the internal structure of planets discovered by transit surveys. 
Finally, we show that due to the strong orbital perturbations, it is possible that the spin of the outer resonant planet of Kepler-1705 is trapped in a sub- or super-synchronous spin--orbit resonance. This would have important consequences for the climate of the planet because a non-synchronous spin implies that the flux of the star is spread over the whole planetary surface.
}


\maketitle

\section{Introduction}

The most successful technique for detecting exoplanets ---in terms of number of planets detected--- is the transit method: when a planet passes in front of a star, the flux received from that star decreases. This technique has been, is being, and will be applied by several space missions such as CoRoT, Kepler/K2, TESS, and the upcoming PLATO mission, to try and detect planets in large areas of the sky. When a single planet orbits a single star, its orbit is periodic, which implies that the transit  happens at a fixed time interval. This constraint
%basic implication from the Keplerian laws 
is used to detect planets when their individual transits are too faint with respect to the noise of the data: using algorithms such as Boxed Least Squares \citep[BLS,][]{Kovacs2002}, the data-reduction pipelines of the transit survey missions fold each light
curve over a large number of different periods and look for transits in the folded data \citep[][]{Jenkins2010,Jenkins2016}. This folding of the light
curve increases the number of observations per phase, and therefore improves the signal-to-noise ratio (S/N) of any transit.

As soon as two or more planets orbit around the same star, their orbits cease to be strictly periodic. In some cases the gravitational interaction of planets can generate relatively short-term transit timing variations (TTVs): transits no longer occur at a fixed period  \citep{Dobrovolskis1996,Agol2005}. The amplitude, frequency, and overall shape of these TTVs depend on the orbital parameters and masses of the planets involved \citep[see e.g.][]{Lithwick2012,NeVo2014,AgolDeck2016}. As the planet--planet interactions that generate the TTVs typically occur on timescales longer than the orbital periods, space missions with longer baselines such as Kepler and PLATO are more likely to observe such effects. Since the end of the Kepler mission, several efforts have been made to estimate the TTVs of the Kepler Objects of Interest (KOIs) \citep{Mazeh2013,RoTho2015,Holczer2016,Kane2019}. 

Transit timing variations are a goldmine for our understanding of planetary systems: they can be used to constrain the existence of non-transiting planets, thereby adding missing pieces to the architecture of the systems \citep{Xie2014,Zhu2018}, and allowing for a better comparison with synthetic planetary system population synthesis models \citep[see e.g.][]{Mordasini2009, Alibert2013,Mordasini2018,Coleman2019,Emsenhuber2020}. TTVs can also be used to constrain the masses of the planets involved \citep[see e.g.][]{Nesvorny2013}, and therefore their density, which ultimately provide constraints on their internal structures, as is the case for the Trappist-1 system \citep{Grimm2018,Agol2020}. Detection of individual dynamically active systems also provides valuable constraints on planetary system formation theory, as the current orbital state of a system can display markers of its evolution \citep[see e.g.][]{BaMo2013,Delisle2017}. Orbital interactions also impact the possible rotation state of the planets \citep{DeCoLeRo2017}, and therefore their atmosphere \citep{Leconte2015}. 

However, TTVs can also prevent the detection of exoplanets. As previously stated, transit surveys rely on stacking the light
curve over a constant period to extract the shallow transits from the noise. If TTVs of  amplitude  comparable\ to or greater than the duration of the transit   occur on a timescale comparable to or shorter than the mission duration, there is not a unique period that will successfully stack the transits of the planet \citep{Garcia2011}. 
%As we will see in section \ref{sec:prob}, t
This can lead to two problems: incorrect estimates of the planet parameters, and/or the absence of detection. In particular, \cite{Kane2019} identified that the lack of known small planets with large TTVs is consistent with an observational bias. In recent years, approaches have been developed to recover the correct transit parameters of known planets exhibiting TTVs, such as for example the photo-dynamical model of the light
curve \cite{RaHo2010}, or the spectral approach by \cite{Ofir2018}. However, these approaches  require detection of the planets. The QATS algorithm alleviates part of this bias \citep{CarAg2013,Kruse2019}, relaxing the strict periodic constraint of other detection algorithms such as the BLS, by allowing the duration between transits to vary within a given range. This gives the algorithm the clear advantage of being able to correct for any TTV shape within a given range of instantaneous periods, but it also leads to a decrease in efficiency for larger TTVs and smaller S/Ns of individual transits, as the stochastic background increases with the range of allowed instantaneous periods.

In recent years, neural networks have been used to vet planetary transits and other astronomical phenomena, or even to detect new planets \citep[see e.g.][]{Shallue2018,Pearson2018,Osborn2020,Armstrong2020}, with approaches that were based on the study of individual transits. In this paper, we present a transit detection method that is adapted to low-S/N planets ---i.e. with individual transits that are shallower than the standard deviation of the photometric flux--- and is robust to TTVs. To do so, we formulate the problem with an image-recognition approach \citep{krizhevsky2012imagenet}.

The paper is structured as follows: in section \ref{sec:prob} we discuss the problem of TTV bias. In Section \ref{sec:RIVERS} we introduce the RIVERS (recognition of interval variations in exoplanet recovery surveys) method using Kepler-36b as an example. In section \ref{sec:detection}, we use the RIVERS method to detect and characterise a pair of resonant planets around Kepler-1705. The dynamics of the resonant pair is discussed in section \ref{sec:dyn}. Finally, we discuss the choices and caveats of the method, and conclude in section \ref{sec:discncon}.

%transit vetting

%transit discovery based of the shape of individual transit. 

%The RIVERS project introduced in this paper explore another approach to recover the signature of small planets with large TTVs. 
 %first and foremost, the predicted yield of this project is based on synthetic population of planetary system \citep{Mordasini2018}. Our yield (after the estimation of our final bias) will hence provide further constrain to compare observations to the theory of formation and evolution of planetary systems.  TTVs provide information on the masses of planets,

%\section{The RIVERS method}

\section{The problem}
\label{sec:prob}

%\subsection{the TTV bias}
 \begin{figure}[!ht]
\begin{center}
\includegraphics[width=0.49\textwidth]{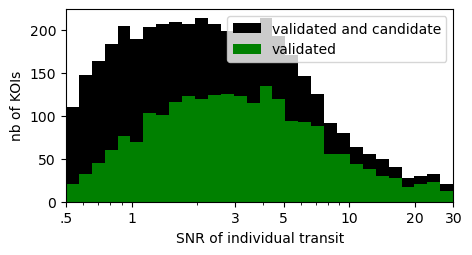}
\includegraphics[width=0.49\textwidth]{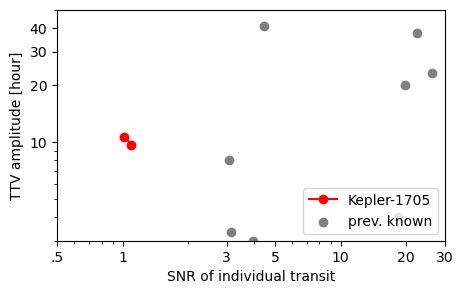}
\caption{\label{fig:kepler} \textit{Top:} Individual transit S/N for all validated KOIs  (Exoplanet Archive Disposition of `confirmed') or candidate (disposition of `candidate') as of May 2021. \textit{Bottom:} Peak-to-peak TTV amplitude of Kepler planets and candidates as a function of the S/N of individual transits, for planets with TTVs of more than 3 hours of peak-to-peak amplitude. Previously known systems include: Kepler-30 \citep{Panichi2018}, Kepler-88 \citep{Nesvorny2013}, KOI-227 \citep{Nesvorny2014}, Kepler-223 \citep{Mills16}, 
%Kepler-9 \citep{Holman2010}, 
Kepler-603 \citep{Holczer2016}, and Kepler-36 \citep{Carter2012}.}
\end{center}
\end{figure}

%The , used in 
As mentioned in Sect. 1, the data-reduction pipeline of the transit survey missions such as Kepler/K2 and TESS \citep{Jenkins2010,Jenkins2016} uses the BLS algorithm  \citep[][]{Kovacs2002}, which folds each light
curve over a large range of different periods and looks for transits in these folded data. 

For a dataset with a flux $f$ associated with the time $t$, folding over the period $P$ is equivalent to mapping $f$ to $\varphi = \text{mod}(t, P)$. Folding the light
curve therefore increases on average the number of measurements per bin of phase $\varphi$ of the chosen period $P$ by a factor $N$, where $N$ is the number of times the light
curve was folded. As a result, the S/N of the transits of the planet in the folded light
curve is increased by a factor $\sqrt{N}$ compared to the individual transits in the unfolded light
curve. %This method allows to detect planets even if individual transit cannot be identified in the noise.

In multi-planetary systems, planet--planet interaction can generate TTVs \citep{Dobrovolskis1996,Agol2005}. In particular, planets near or in mean motion resonance\footnote{Two-body MMRs are defined by $P_2/P_1\simeq (k+q)/k$, where $k$ and $q$ are integers, while three-body resonances are defined by $(k+q)/P_2 \simeq k/P_1+q/P_3$} (MMR) can exhibit significant TTV over a timescale comparable to or smaller than the baseline of transit surveys such as Kepler and PLATO. The typical timescale of these configurations and their dependence on the planetary masses and eccentricities are given in Appendix \ref{sec:TTV}.

We define S/N$_i$ as the S/N of an individual transit of a planet:
\begin{equation}
\text{S/N}_i = \sqrt{N_\text{transit}} D/\sigma_{lc}   \, ,
\label{eq:Depth}
\end{equation}
where $D$ is the depth of the transit, $N_\text{transit}$ is the number of measurements during one transit and $\sigma_{lc}$ is the standard deviation of the flux of the light
curve. When these configurations induce TTVs that have a period comparable to or shorter than the duration of the data and a peak-to-peak amplitude $\sigma_{TTV}$ comparable to or larger than the transit duration $T_{transit}$ of the planet, the smearing of the folded transit leads to alteration of the determined transit parameters. In addition, the S/N of the transit is reduced, which can prevent the detection of the planet.
%by a factor $\approx (T_{transit}/\sigma_{TTV})^{1/2}$ \citep{Garcia2011,CarAg2013}. 
Assuming the TTVs are sinusoidal and the transit is box-shaped, for observations longer than the TTV period the depth of the stacked transit can be estimated by (see Appendix \ref{ap:smearing}):
\begin{equation}
\text{Smeared Depth} = D/(1+\sigma_{TTV}/ T_{transit})  \, .
\label{eq:Depth}
\end{equation}
 We can therefore estimate the smeared S/N of a planet using the following formula:
 %np.sqrt((1+2*TTVsd)/(1+4*TTVsd**2+4*TTVsd+2*TTVsd/siglc**2)
\begin{equation}
\text{Smeared S/N} = \frac{\text{S/N}_i}{1+\sigma_{TTV}/ T_{transit}} \sqrt{ \frac{\text{Mission duration}}{\text{Orbital period}}}
\label{eq:TTVsec}
.\end{equation}

In the Kepler mission, if the smeared S/N is below 7.1, the signal will not be considered as a KOI. This is probably why all known Kepler planets with a large TTV also have a large S/N$_i$. The top part of Fig. \ref{fig:kepler} shows the S/N$_i$ of all KOIs. The bottom part shows the known planets with a peak-to-peak TTV of greater than 3 hours and a S/N$_i$ below 30. All of these planets have a S/N$_i$ above 3, which allowed either their detection in a stacked light
curve despite the smearing of the transit or the detection of their individual transits. The comparison of the two figures shows an over-representation of planets with large S/N$_i$ in the population of planets with large TTVs. If anything, the underlying expected trend is the opposite: for a pair of planets in or near mean-motion resonance, the relative amplitude of TTVs between the two planets is proportional to $TTV_1/TTV_2 \propto m_2/m_1$, yielding larger TTVs for the least massive planet of the pair. In addition, in the co-orbital resonance, the lower the sum of the masses of the two planets, the larger the TTV amplitude can be \citep{Leleu2019}. We therefore expect a subpopulation of planets with large TTVs and small S/N$_i$ that is currently missed in transit surveys because of the TTV bias.

\section{RIVERS approach}
\label{sec:RIVERS}

We aim to alleviate the bias that prevents standard algorithms from recovering and characterising the small, dynamically active planets that could have been missed in transit surveys. To do so, we need to recover in a given light
curve a large number of transits that are individually too small to be identified. We therefore skip the determination of individual transits and focus on the fit of the light
curve by transit models whose timing is constrained using TTV models. The model ensures that the TTVs follow a signal that is physically possible, reducing the space of available TTVs compared to an approach in which each transit timing is freely fitted to the data. This approach also reduces the number of free parameters to a maximum of five per planet for a coplanar system, against one free parameter per transit. The downside is that the method will only be able to identify planets whose TTVs are dominated by the effect of the planets that are considered in the model. The TTVs can be modelled by N-body simulations, allowing for an accurate TTV prediction regardless of the number of modelled planets and orbital configuration \citep[see e.g.][]{DeAgHo2014}. In the case of two interacting planets,  analytical models can be used for specific
configurations; see for example \cite{AgolDeck2016} and \cite{DeckAgol2016} for the TTVs of a pair of planets outside of MMR, and \cite{NeVo2016} for the TTVs of a pair of planets in first-order MMR. The analytical approach can be up to two orders of magnitude faster \citep{DeckAgol2016} and further reduce the number of free parameters.

This approach uses TTV modelling as a means of detection in addition to a tool for the characterisation of the system. However, these fits are too time consuming to be applied to all Kepler stars for any orbital period. We therefore need an algorithm to identify promising orbital periods for a given star which does not rely on the exact periodicity between the transits to detect candidates. We base our approach on shape recognition in a 2D representation of the light
curve called a river diagram \citep[or riverplot, introduced by][]{Carter2012}. We illustrate this method in the following section using the well-known example of Kepler-36b.

\subsection{ Kepler-36b}

 \begin{figure}[!ht]
\begin{center}
\includegraphics[width=0.49\textwidth]{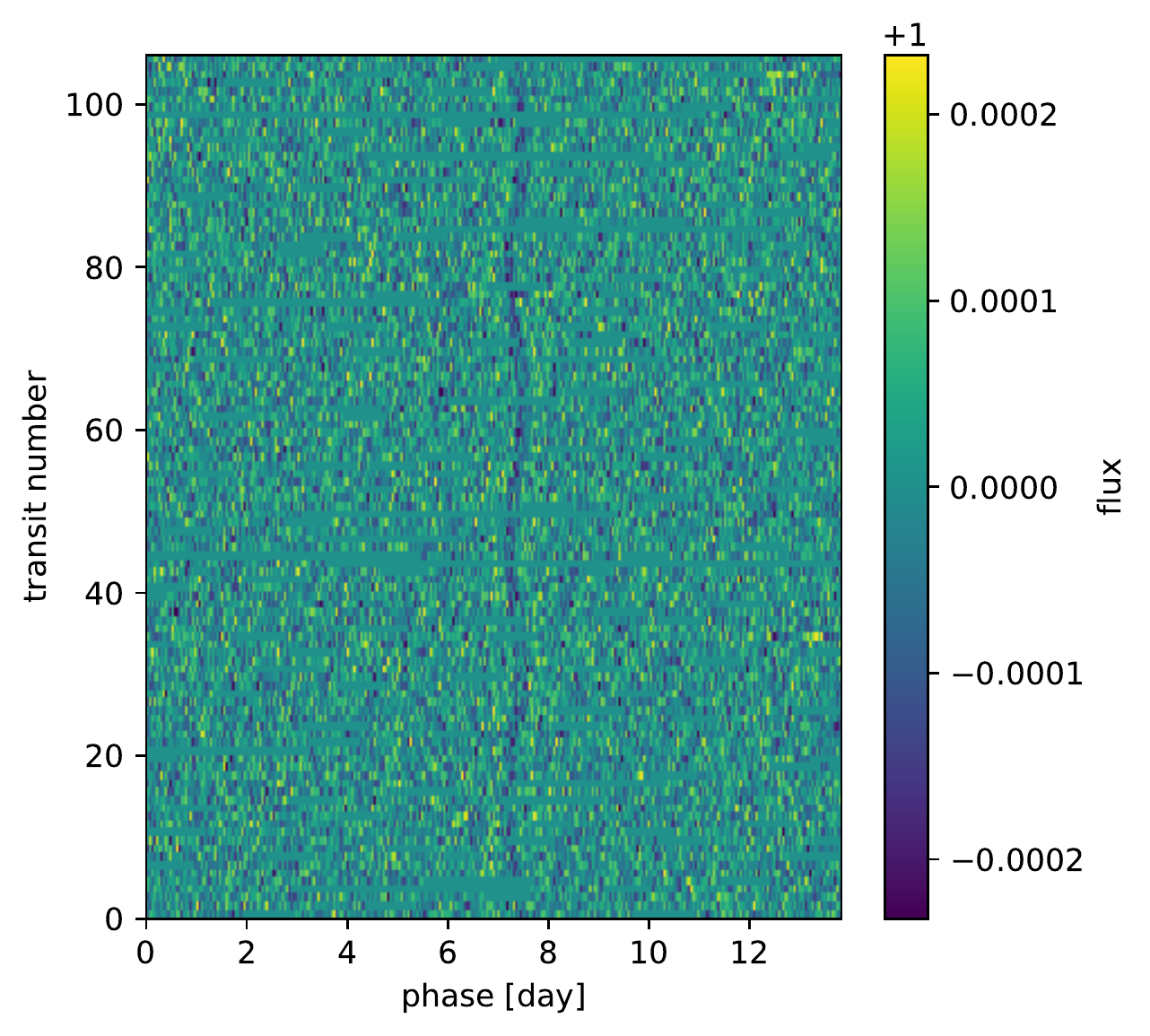}
\caption{\label{fig:river_Kepler36} River diagram of Kepler-36 at the period 13.848d: the bottom row displays the first 13.848 days of data for Kepler-36, with the colour code representing the normalised flux. Each subsequent row displays a new set of 13.848 days of data. The flux has been clipped at $3\sigma$ for visibility, and missing data have been replaced by a flux of 1.}
\end{center}
\end{figure}

Kepler-36b has a TTV amplitude of about 8 hours for a transit duration of 7 hours \citep{Carter2012}. The river diagram of the DPCSAP flux of Kepler-36 for  $P_\text{fold}=13.8480\,$days is shown in Fig. \ref{fig:river_Kepler36}. In a river diagram, each line displays the normalised flux coming from the star over a single (constant) orbital period $P_\text{fold}$. When $P_\text{fold}$ is close to the average orbital period of a planet or its aliases, subsequent transits of the planets vertically align, allowing the eye to track the signature of the planet. A planet without TTVs produces a straight line, while TTVs induce variations in the horizontal position of each transit. If $P_\text{fold} \approx P_\text{orb}$, a single curve appears. If $P_\text{fold} \approx k P_\text{orb}$, with $k$ an integer, $k$ full curves appear in the diagram. More generally, if $j P_\text{fold} \approx k P_\text{orb}$ with $k$ and $j$ being positive integers, $k$ curves appear, with a transit every $j$ lines.

\begin{figure}[!ht]
\begin{center}
\includegraphics[width=0.49\textwidth]{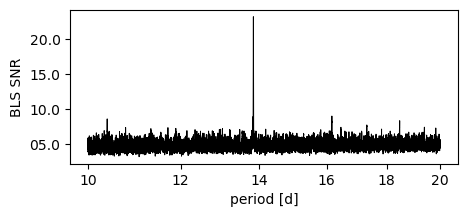}

\includegraphics[width=0.49\textwidth]{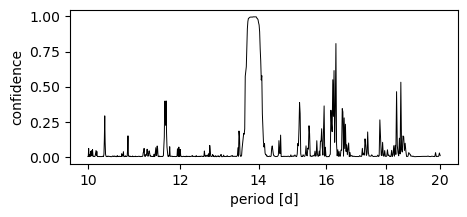}
\caption{\label{fig:RIVERSperiodo_Kepler36} Periodograms of the PDCSAP flux of Kepler-36, once the transits of Kepler-36c (16.23d) are masked. \textit{Top:} BLS periodogram. \textit{Bottom:} RIVERS periodogram. The y axis represents the confidence in the model that the river diagram folded at the period shown on the x-axis has a planet-like track.  }
\end{center}
\end{figure}

The track of Kepler-36b, with an S/N$_i$ of $3.07$, is clearly visible in the river diagram. The S/N$_i$ is sufficient to recover the planet in the BLS despite the smearing: the top panel of Fig. \ref{fig:RIVERSperiodo_Kepler36} shows the BLS periodogram applied to the the pre-search data-conditioning simple aperture photometry (PDCSAP) flux of Kepler-36, once the transits of Kepler-36c (16.23d) are masked. An analysis of the light
curve using the QATS algorithm was able to recover the individual transits, leading to characterisation of the resonant pair Kepler-36b and  Kepler-36c \citep{Carter2012}.

\subsection{The RIVERS.deep approach}
\label{sec:approach}

\subsubsection{Model architecture}
\label{sec:model}

\begin{figure}
  \begin{center}
     \includegraphics[width = 9cm]{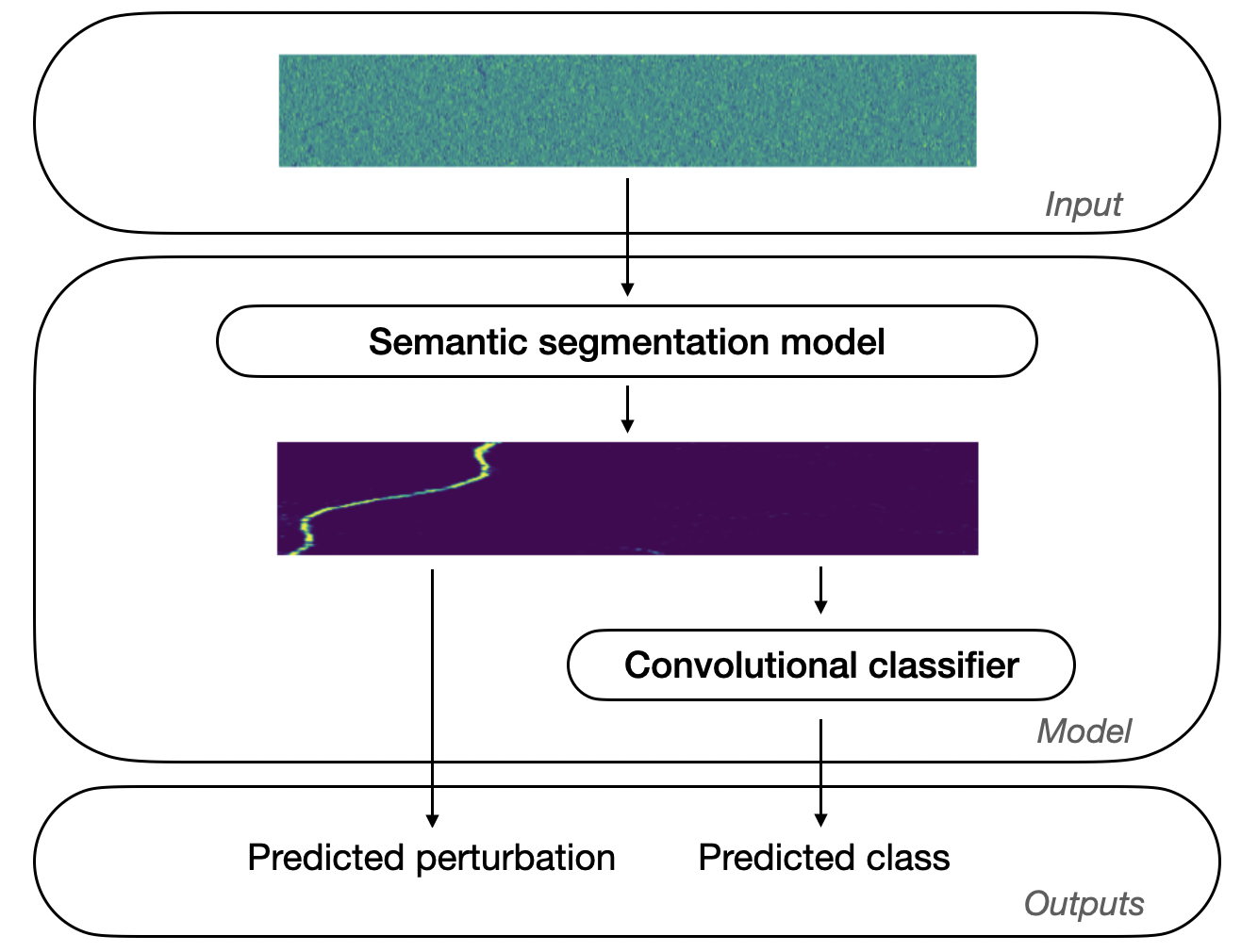}
  \end{center}
  \caption{Neural network model architecture. The model takes a river
    diagram as input, computes a predictions mask, and feeds it into a
    convolutional neural network (CNN) classifier. The model output is
    the couple (predicted mask, CNN prediction).}
  \label{fig:arch}
\end{figure}

As long as $P_\text{fold}$ is close enough to the average orbital period of the planet, river diagrams are a 2D representation of the light
curve in which the track of a planet draws a curve. As neural networks have been shown to be well suited to performing pattern recognition \citep{krizhevsky2012imagenet}, we developed the RIVERS.deep algorithm, which takes as input a river diagram such as that shown in Fig. \ref{fig:river_Kepler36} and produces two outputs:
\begin{itemize}
\item {\bf A confidence matrix:} An array of the same size as the input containing for each pixel the confidence that this pixel belongs to a transit. This task is performed by the `semantic segmentation' (pixel-level vetting) subnetwork \citep{jegou2017one}. 
\item {\bf global prediction:} A value between 0 and 1 which quantifies the model confidence that the output of the semantic segmentation module is due to the presence of a planet. This task is performed by the classification subnetwork.
\end{itemize}

Our model works with arrays of  fixed size, while the  sizes of river diagrams depend on the duration of the dataset and $P_\text{fold}$ for a fixed bin size (taken at $30\,$min for this study). The river diagrams are therefore all resized to a given size using the nearest interpolation method, which gave the best results amongst the methods  tested (bilinear interpolation, nearest, padding the missing pixels with values of 1, and a mosaic repetition of the river diagram). To avoid losing information, the number of pixels along the x-axis is determined by the longest considered period divided by the bin size, while the y-axis is set by the duration of the observations divided by the shortest considered period. To avoid too much stretching of the matrices, we trained different models for different period ranges: 5 to 10 days, 10 to 20 days, and 20 to 30 days. The global architecture of the model is illustrated in Fig.~\ref{fig:arch} and a precise description of the components is available in Appendix~\ref{ap:model}. Several architecture experiments are also discussed in that Appendix.

\subsubsection{Training set}
\label{sec:training_set}  

In order to train the neural network, we generated a large set of inputs (river diagrams), and their expected outputs: a transit mask and a boolean indicating the presence of a planet with a mean period close to the folded period. Transit masks are arrays of the same shape as the river diagram. If the river diagram contains the track of a planet, the transit mask contains `1' in transits and  `0' everywhere else. When the river diagram does not contain a planet, the transit mask is full of  zeros. For best performance of the model, the training dataset has to be as close as possible to the real data; it is therefore preferable to train a model dedicated to a given mission. In this paper, we focus on the Kepler light
curves as their duration allows us to use exploit the full potential of our method with its 4 years of continuous observation. Ideally, we would train our model on real planets. However there are not enough planets with large TTVs and low S/N to constitute a training set. 

We therefore generated a synthetic training set of 40,000 river diagrams for each of the period ranges mentioned in the previous section. To do so, we generated a set of 40,000 unique orbital configurations: to train a model to recognise planets with periods in the range of 10 to 20 days, the period of the transiting planet $P_1$ is randomly chosen between $10.2$ and $19.8\,$d with a flat probability. The (neighbouring) MMR is then randomly picked amongst $k:k+1$ or $k+1:k$ for $1 \leq k \leq 8$,  $k:k+2$ or $k+2:k$ for $k=\{1,3,5\}$, or the $1:1$ MMR. The initial period of the non-transiting planet is taken at the exact resonance multiplied by $1+3 \theta  \sqrt{(m_1+m_2)/m_\star}$, where $\theta$ is randomly chosen between $-0.5$ and $0.5$. The eccentricities are chosen randomly between $0$ and $0.1$, and all angles are randomised. The systems were integrated as coplanar, although only planet 1 is injected in the light
curve, which should not significantly impact the results, as analytical studies point out that small mutual inclinations do not impact the shapes of TTVs \citep{NeVo2016}. Only systems that were short-term unstable (collision or ejection from the initialised configuration over the 10 years of integration) were removed from this dataset.

Each of these 40,000 orbital solutions is then injected in one of the selected Kepler light
curve. The light
curves were selected through the {\ttfamily kplr} package, with the following filters: The light
curve has to contain at least one {\ttfamily confirmed} planet of period between 3 and 50 days, and no {\ttfamily false\! positive} or {\ttfamily not\! dispositioned} KOIs, regardless of their period.
%In addition, lightcurves with known planets with TTV amplitudes of more than three hours were removed as well. 
This resulted in about 1,200 unique light
curves.
%we then inject the signal of of the planet of orbital period $P_1$ into a Kepler lightcurve. 
%
%To do so, we inject the signal of a planet with large TTVs into a lightcurve from the Kepler mission. 
For each light
curve, the raw PDCSAP flux is downloaded using the {\ttfamily lightkurve}\footnote{https://docs.lightkurve.org/} package. The signal of all known planets or KOIs is masked: data points falling in or near the transits are removed from the dataset. For planets with known TTVs, we removed an additional width of 1.5 times the TTV amplitude reported in \cite{Kane2019}. 
%In parallel to this treatment of the lightcurve, the trajectory of a two-planets system including a $10\, m_{\oplus}$ on a $18.1$ day orbit and a $30\, m_{\oplus}$ on a $25.15$ day orbit around a sun-like star is integrated over 5 years. This system, close to the $7/5$ MMR exhibit significant TTVs (see Fig. \ref{fig:example_TTV}). Assuming that only the inner planet is transiting, 

To inject the signal of the perturbed planet in a Kepler light
curve, its effect on an ideal normalised light curve (noiseless) is computed at the date of each data point of the light curve using the {\ttfamily batman} package \citep{batman}. The parameters of the star, such as its radius and limb-darkening coefficients, are obtained through the {\ttfamily kplr}\footnote{http://dfm.io/kplr/} package. The raw PDCSAP flux and the idealised light curve are then multiplied data point by data point, simulating the transit of the injected planet. The obtained light curve is then detrended using the \textit{flatten}\footnote{https://docs.lightkurve.org/reference/api/lightkurve.LightCurve.flatten.html} method of the {\ttfamily lightkurve} package using the default parameters. This method applies a Savitzky-Golay filter to the light curve. The whole process results in a synthetic light curve containing the transits of a dynamically active planet, with noise structure identical to that particular Kepler target.

Finally, we split the light curves into two groups. With half of the synthetic light curves we generate a river diagram with $P_{fold}$ close enough to the average period $P$ of the injected planet so that the track of the planet appears only once per line in the diagram. For a planet without TTVs, this corresponds to a range of periods in which the track of the planet makes anything from a bottom-left to top-right diagonal of the river diagram, to a bottom-right to top-left diagonal, hence $P- \delta P <P_{fold} <P+\delta P$, where 
\be
\delta P = P^2/T \, ,
\label{eq:dP}
\ee

 where T is the mission duration. These are the river diagrams we label as {\ttfamily goodfold} and correspond to the presence of a planet for the classifier subnetwork. We do not train only on diagrams where $P_{fold}=P$ because $P$ is not known in advance when looking for new planets, and therefore the model needs to be able to recognise tilted planet tracks. The remaining halves of the light curves are used to generate river diagrams with a random period in the range of 10 to 20 days, excluding periods that correspond to $P- 2 \delta P <P_{fold} <P+ 2 \delta P$. We label those as {\ttfamily badfold}, corresponding to the absence of a planet at $P_{fold}$ for the classifier subnetwork.

\subsubsection{Periodogram}

In order to visualise whether or not a given star harbours promising candidates, we generate a periodogram of the light curve by creating river diagrams for a subset of $P_{fold}$ that span the range of periods for which the model was trained. As discussed in the previous section, we need to ensure that regardless of the period $P$ of the potential planet, there are river diagrams that will be evaluated at $P_{fold}$ such that $|P_{fold}-P| <  \delta P$ (Eq. \ref{eq:dP}). As the recovery rate of the model is not $100\%$, we sample the period range with a smaller period step by introducing the parameter $N_{diag}$.
% which represent the number of frame in which the track of a planet with period $P$ will appear between a diagonal from the top left corner of the river diagram to the bottom right
Starting at $P_0$, which is the smallest period on which the model is trained, the set of periods $P_n$ that form the periodogram is defined recurrently, following:
\be
 P_n=P_{n-1} + \delta P_{n-1}/  N_{diag}\, .
\ee
For each $P_n$, the periodogram shows the confidence of the model that the river diagram contains the track of a planet.

\subsubsection{Application of the model}

\begin{figure}[h!]
\begin{center}
\includegraphics[width=0.49\textwidth]{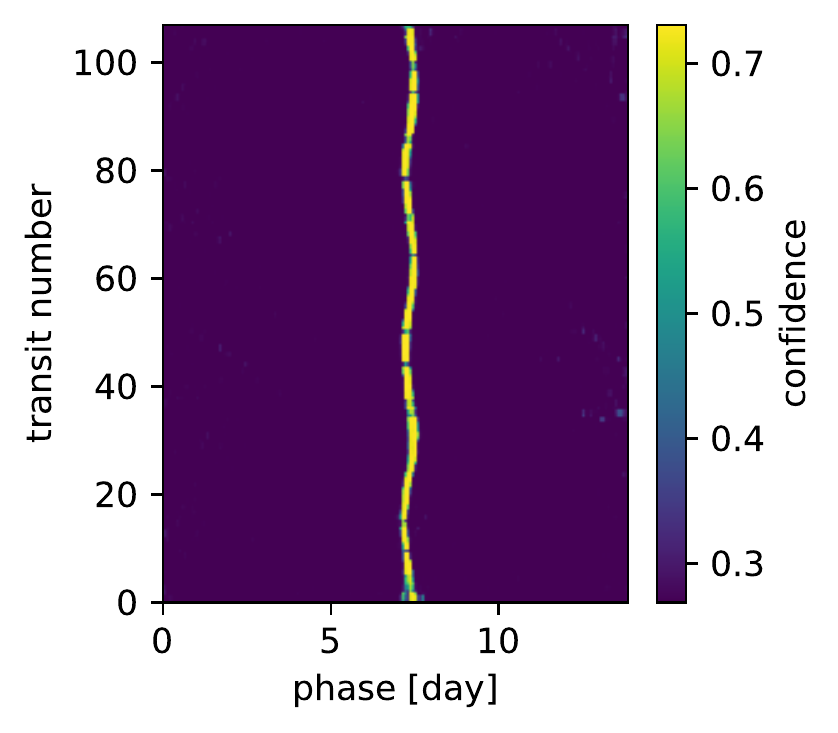}
\caption{\label{fig:RIVERSdeep_Kepler36} RIVERS.deep confidence matrix. Each pixel corresponds to the same bin of the light curve as the corresponding river diagram shown in Fig. \ref{fig:river_Kepler36}. The colours show the confidence of the model that the timing belongs to the track of a planet.  }
\end{center}
\end{figure}

The bottom panel of Fig. \ref{fig:RIVERSperiodo_Kepler36} shows the RIVERS.deep periodogram applied on the PDCSAP flux of Kepler-36, once the transits of Kepler-36c (16.23d) are masked. A wide peak of confidence of approximately $1$ appears near $P_\text{fold}=13.8 \,$days which is the average orbital period of Kepler-36b. The river diagram shown in Fig. \ref{fig:river_Kepler36} belongs to the peak of this periodogram. Figure \ref{fig:RIVERSdeep_Kepler36} shows the confidence matrix corresponding to the same river diagram, where the semantic segmentation task highlighted the timings that belong to the transits. These timings can be used to estimate a proxy of transit timings that can then be used to perform a preliminary fit of the TTV model, ensuring that the subsequent fit of the light curve is initialised near the solution. In addition, stacking the light curve along the proxy of the transit timings allows us to approximate the transit parameters such as the planetary radii and impact parameter for the initialisation of the photodynamic fit. In the following section, we show how this approach allowed us to discover and confirm a pair of resonant planets around  Kepler-1705.

 %Note that the peak has a non-zero width: since we do not know \textit{a-priori} the period of the planet, the model was trained to recognised not only vertical structures (where $P_\text{fold}$ is exactly the average period of the planet), but tilted ones as well.

\section{Detection and study of Kepler-1705b and Kepler-1705c}
\label{sec:detection}

 \begin{figure}[h!]
\begin{center}
\includegraphics[width=0.49\textwidth]{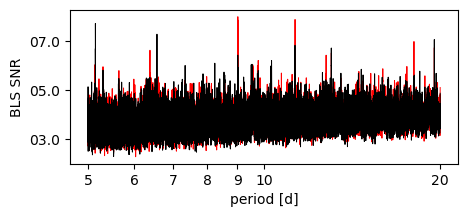}
\includegraphics[width=0.49\textwidth]{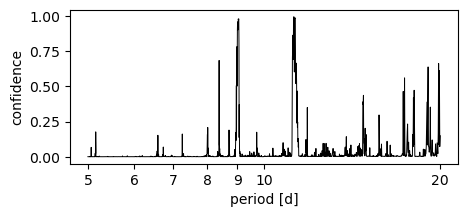}
\caption{\label{fig:RIVERS_periodo} Periodogram of the light curve of Kepler-1705 after removal of the 3.08d candidate. \textit{Top:} BLS periodogram on the first 12 quarters in red, on the full dataset in black. \textit{Bottom:} RIVERS periodogram on the full dataset. The peaks around 9.03d and 11.3d have a non-zero width, implying that subsequent frames had a high confidence from the model. Examples of river diagrams for each of these peaks are given in Fig. \ref{fig:RIVERS9p0}.}
\end{center}
\end{figure}

Kepler-1705 (KOI4772, KIC 8397947) is an F star (6300K) with a magnitude of $mV= 15.800\pm0.206$, $m_{Kep}=15.514$. As of October 5, 2021, two candidates at 3.38\,d (KOI4772.01) and 9.01\,d (KOI4772.02) and a false positive at 39.09\,d (KOI4772.03) are announced on the Kepler database\footnote{\href{https://exoplanetarchive.ipac.caltech.edu/cgi-bin/TblView/nph-tblView?app=ExoTbls\&config=cumulative}{\url{https://exoplanetarchive.ipac.caltech.edu/cgi-bin/TblView/nph-tblView?app=ExoTbls\&config=cumulative}}}. An analysis by the BLS algorithm of the flattened PDCSAP flux (section \ref{sec:training_set}) of all available quarters allows  KOI4772.01 to be recovered at the announced 3.38d period. %, see top panel of Fig. \ref{fig:BLS}.
However, once the transits of KOI4772.01 are masked, a second use of the BLS yields no significant peak in the range of 5 to 20 days; see the black periodogram in the top panel of Fig. \ref{fig:RIVERS_periodo}.

\subsection{Application of the RIVERS.deep method} 

 \begin{figure*}[h!]
\begin{center}
\includegraphics[width=0.49\textwidth]{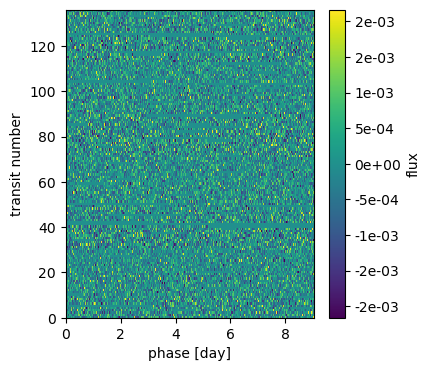}\includegraphics[width=0.49\textwidth]{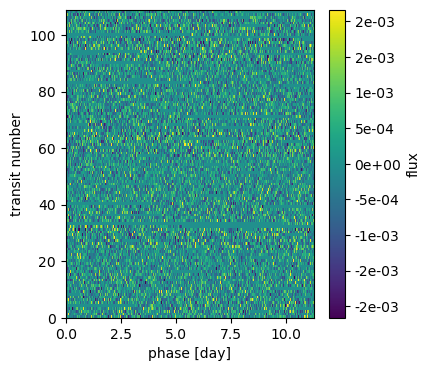}

\includegraphics[width=0.49\textwidth]{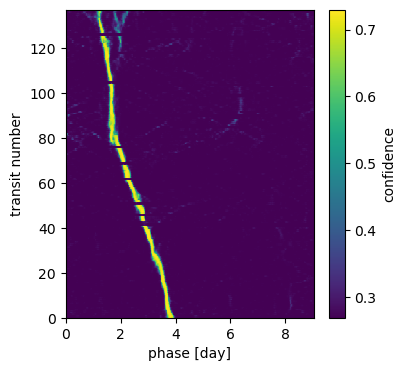}\includegraphics[width=0.49\textwidth]{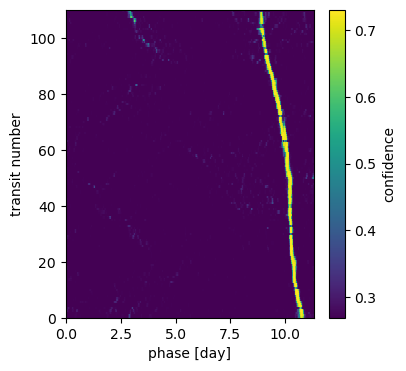}
\caption{\label{fig:RIVERS9p0} \textit{Top:} River diagrams of Kepler-1705 at the period of 9.0471d (\textit{left}) and 11.3059 (\textit{right}). The bottom left corner starts at 352.3975 [BJD-2454833.0] for both. See Fig \ref{fig:river_Kepler36} for more details.
% : the bottom raw displays the first 9.0471 days of data for KOI4772, the color code representing the normalised flux. Each subsequent raw display a new set of 9.0471 days of data. The flux has been clipped at $3\sigma$ for visibility. 
\textit{Bottom:} Corresponding RIVERS.deep confidence matrices, which shows  the confidence for each timing of the rivers diagram to belong to the track of a planet. See section \ref{sec:model} for more details about the confidence matrix.}
\end{center}
\end{figure*}

 \begin{figure}[h!]
\begin{center}
\includegraphics[width=0.49\textwidth]{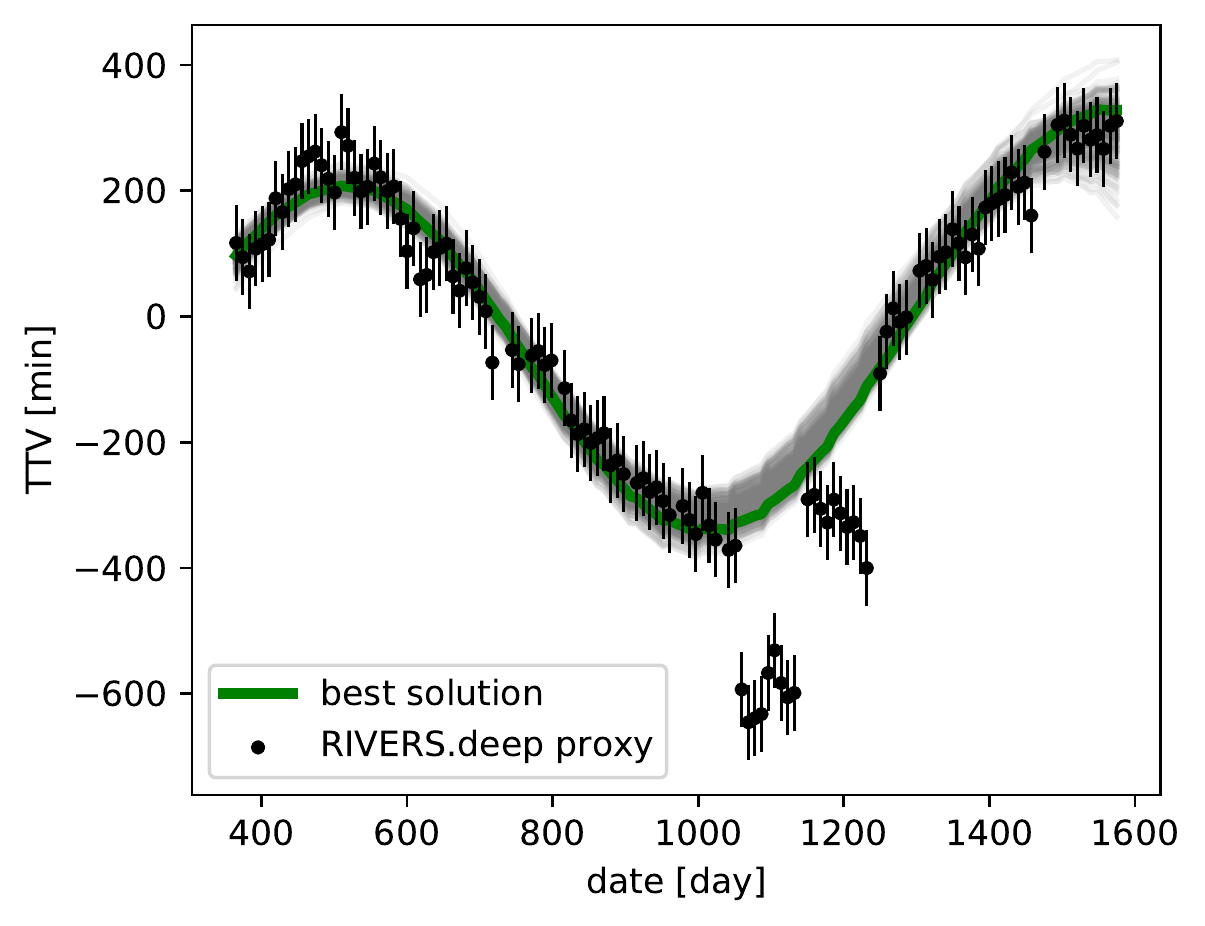}
\includegraphics[width=0.49\textwidth]{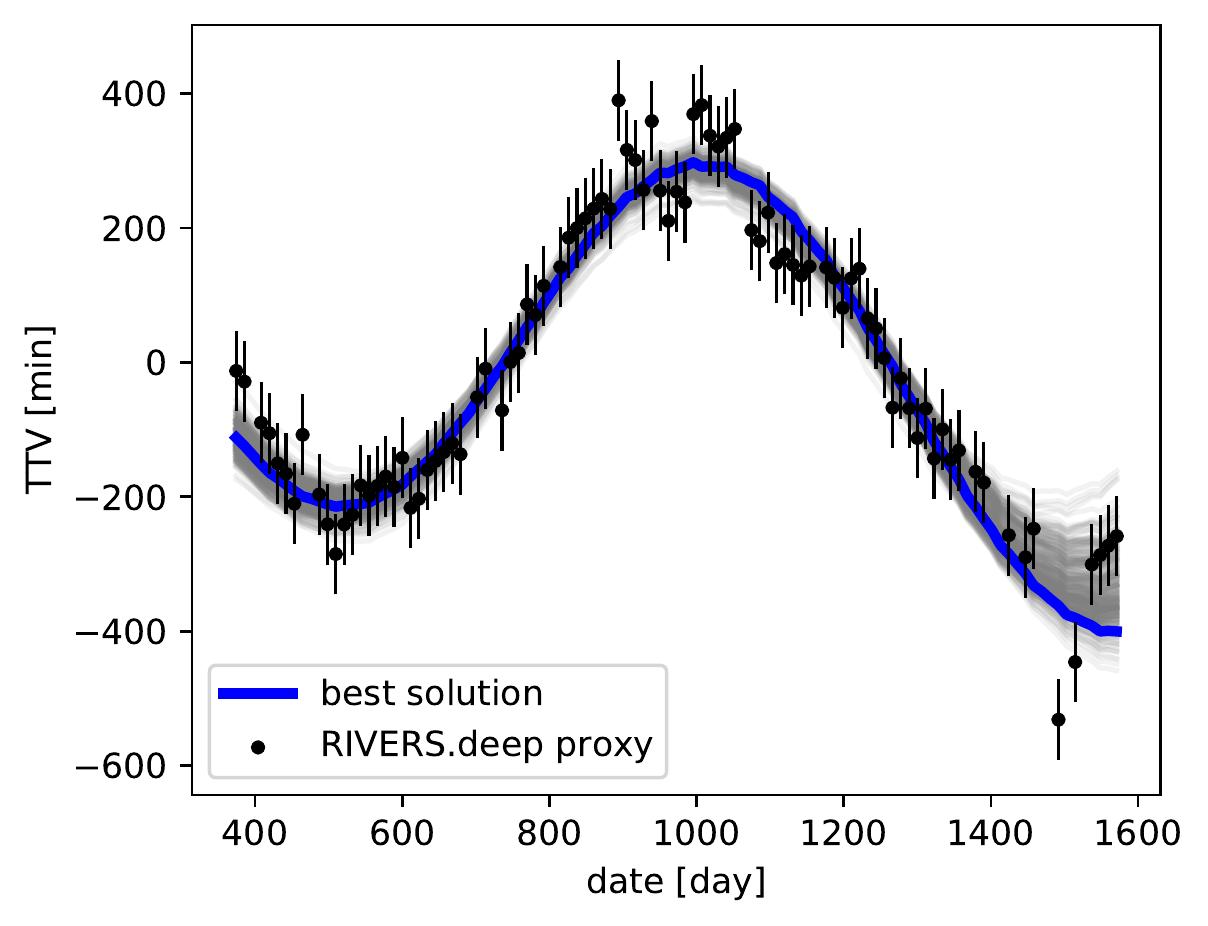}
\caption{\label{fig:KOI4772_TTV} TTVs for Kepler-1705b (\textit{top}) and Kepler-1705c (\textit{bottom}). The black error bars represent the TTVs proxy coming from the RIVERS.deep method. The dates are in [BJD-2454833.0]. In grey are 300 samples resulting from the fit of the light curve. The solid coloured curves correspond to the best fit.}
\end{center}
\end{figure}

 We further analysed the light curve using the approach explained in Section \ref{sec:RIVERS}. The bottom panel of Fig. \ref{fig:RIVERS_periodo} shows the RIVERS periodogram applied on the same dataset as the black BLS periodogram shown in the top panel. In the RIVERS periodogram, two peaks appear at $\sim 9.0$ and $\sim 13.2$ days. The river diagram and RIVERS.deep matrix for a frame chosen in each of these peaks are shown in Fig. \ref{fig:RIVERS9p0}. For each of these periods, the model identified a coherent track for the entire duration of the Kepler mission. 
We then retrieved the proxy of transit timings from each of the RIVERS.deep diagrams of Fig. \ref{fig:RIVERS9p0}. The TTVs corresponding to these transit timings are shown with black circles in Fig. \ref{fig:KOI4772_TTV}. We note that the error bars are simply an indication of the resolution of the river diagram (bins of 30 mins). The two signals appear to be in phase opposition, as expected for the TTVs of two strongly interacting planets. Indeed, the two peaks lie close to the 5:4 MMR. Although caveats of the method are discussed in section \ref{sec:caveats}, we stress that these are not equivalent to transit timings that could have been fitted to the light curve, but the output of a neural network trained to recover the track of TTV-perturbed planets in river diagrams. The validation of the candidates comes from the fit of the light curve.

\subsubsection{Comparison to the result of the BLS}

 \begin{figure}[h!]
\begin{center}
\includegraphics[width=0.49\textwidth]{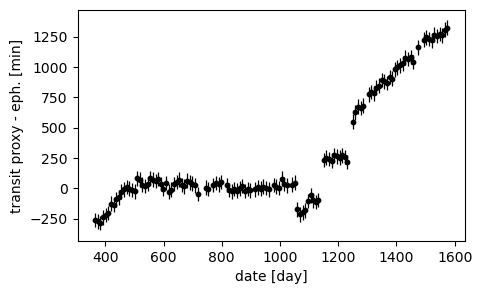}
\caption{\label{fig:RIVBLS} Difference between the RIVERS.deep transit proxy of the RIVERS candidate at $\sim 9$d (Kepler-1705b) and the ephemeride of KOI4772.02 from the exoplanet archive. The dates are in [BJD-2454833.0].}
\end{center}
\end{figure}

As shown in Fig. \ref{fig:RIVBLS}, the timings of the RIVERS candidate at $\sim 9$d (Kepler-1705b) are coherent with the KOI4772.02 candidate published on the exoplanet archive. In the first half of the data, a significant part of the transit timings can indeed be modelled by a straight line, which explains why KOI4772.02 is only found in the Q1-Q12 table of the exoplanet archive: the corresponding peaks appear in the BLS only when analysing the 1-12 quarters; see the red periodogram in the top panel of Fig. \ref{fig:RIVERS_periodo}. On that periodogram, we can see that the 11.3d signal was also on the threshold to become a KOI. This demonstrates that in the BLS periodogram, a signal with TTV can appear with a higher S/N with less data because a locally constant period prevents the smearing of the stacked transit, as discussed in Section \ref{sec:prob}.

\subsection{Planet detection} 
\label{sec:fit}
\begin{table} [h!]
\caption{Stellar properties of Kepler-1705} 
\label{tab:stellarParam} 
\centering 
\begin{tabular}{lll} 
\hline\hline 
\multicolumn{3}{c}{Kepler-1705}\\ 
KIC & \multicolumn{2}{l}{8397947}\\ 
\hline 
Parameter & Value & Note \\ 
\hline 
mKep [mag] & 15.514 & 1 \\ 
mV [mag] & $15.800\pm 0.206$ & 1 \\ 
mJ [mag] & $14.210\pm 0.029$ & 1 \\ 
mH [mag] &  $13.868\pm .031$ & 1 \\ 
\hline 
$T_{\mathrm{eff}}$ [K] & $6312_{-152}^{+215}$ & 2 \\ 
 $\log{g}$ [cgs]      &  $4.298_{-0.047}^{+0.043}$ & 2 \\\relax 
 [Fe/H] [dex] & $-0.20_{-0.15}^{+0.15}$ & 2 \\\relax 
$R_{\star}$ [$R_{\odot}$] &  $1.259_{-0.051}^{+0.032}$  & 2 \\\relax 
$M_{\star}$ [$M_{\odot}$] &  $1.139_{-0.081}^{+0.087}$  & 2 \\ 
$t_{\star}$ [Gyr]  &  $2.59_{-1.61}^{+1.94}$  & 2 \\ 
$L_{\star}$ [$L_{\odot}$] &  $2.26_{-0.24}^{+0.32}$ & 2 \\ 
$\rho_{\star}$ [$\rho_\odot$] &  $0.573_{-0.070}^{+0.077}$  & 2 \\ 
\hline 
\end{tabular} 
\tablefoot{  
 [1] https://exoplanetarchive.ipac.caltech.edu, [2] \cite{Berger2020} } 
\end{table}

\begin{table*} 
\caption{Fitted and derived properties of Kepler-1705b and Kepler-1705c} 
\label{tab:planetParam} 
\centering 
\begin{tabular}{lllll} 
\hline 
Parameter &  Prior & Kepler-1705b & Kepler-1705c  & \\ 
\hline\hline 
$\lambda$ [deg]&U[0,360]&$10.10_{-3.31}^{+4.85}$&$94.75_{-3.06}^{+4.32}$&fitted\\ 
$P$ [day]&U[8,12]&$9.03502_{-6.4e-04}^{+6.5e-04}$&$11.2800_{-0.0009}^{+0.0011}$&fitted\\ 
$e\cos \varpi$ &U*&$0.007_{-0.042}^{+0.029}$&$-0.010_{-0.038}^{+0.026}$&fitted\\ 
$e\sin \varpi$ &U*&$0.010_{-0.024}^{+0.035}$&$-0.003_{-0.022}^{+0.031}$&fitted\\ 
$M_{pl} [M_\star]$ &U[0,1e-2]&$1.2e-05_{-7.6e-07}^{+8.9e-07}$&$1.4e-05_{-1.1e-06}^{+1.2e-06}$&fitted\\ 
$R_{pl} [R_\star]$ &U[0,1e-1]&$0.01479_{-8.2e-04}^{+8.3e-04}$&$0.01491_{-9.1e-04}^{+9.6e-04}$&fitted\\ 
$b$&U[0,1]&$0.32_{-0.20}^{+0.17}$&$0.34_{-0.21}^{+0.19}$&fitted\\ 
$t0$ [BJD-2454833.0]& &$365.214_{-0.011}^{+0.010}$&$374.396_{-0.017}^{+0.020}$&derived\\ 
$a/R_\star$ & &$15.21_{-0.46}^{+0.43}$&$17.64_{-0.53}^{+0.50}$&derived\\ 
$e$ & &$0.033_{-0.020}^{+0.073}$&$0.028_{-0.018}^{+0.064}$&derived\\ 
$\varpi$ [deg]& &$35_{-112}^{+83}$&$-38_{-112}^{+175}$&derived\\ 
$I$ [deg]& &$88.79_{-0.66}^{+0.76}$&$88.89_{-0.70}^{+0.69}$&derived\\ 
$M_{pl}$ [M$_{Earth}$]& &$4.47_{-0.43}^{+0.48}$&$5.42_{-0.57}^{+0.61}$&derived\\ 
$R_{pl}$ [R$_{Earth}$]& &$2.03_{-0.14}^{+0.12}$&$2.05_{-0.15}^{+0.14}$&derived\\ 
$\rho$ [$\rho_{Earth}$]& &$0.54_{-0.10}^{+0.12}$&$0.64_{-0.12}^{+0.15}$&derived\\ 
S/N& &$9.55$&$8.84$& derived \\ 
\hline 
 & & \multicolumn{2}{c}{Kepler-1705} &\\ 
\hline 
\hline 
$\rho_{\star}$ [$\rho_\odot$]&G(0.573,0.077)&\multicolumn{2}{c}{$0.565_{-0.072}^{+0.073}$}&fitted\\ 
limbdark $u_1$&G(0.344,0.035)&\multicolumn{2}{c}{$0.343_{-0.035}^{+0.035}$}&fitted\\ 
limbdark $u_2$&G(0.297,0.043)&\multicolumn{2}{c}{$0.296_{-0.041}^{+0.040}$}&fitted\\ 
log10(jitter)&U[-9,0]&\multicolumn{2}{c}{$-3.603_{-0.011}^{+0.010}$}&fitted\\ 
$log10(\sigma_{GP})$ &U[-9,0]&\multicolumn{2}{c}{$-7.73_{-1.53}^{+1.56}$}&fitted\\ 
$log10(\tau_{GP})$ [day] &U[0,100]&\multicolumn{2}{c}{$1.51_{-1.02}^{+1.01}$}&fitted\\ 
\end{tabular} 
\tablefoot{  
$\lambda$ (mean longitude), $P$ (period), $e$ (eccentricity), and $\varpi$ (longitude of the periastron) are given at the date 2455196.2294 [BJD]. We computed the detection S/N as the median depth of transit over its standard deviation. U(x,y) priors are flat between x and y, G(x,y) priors are Gaussian of expected value x and standard deviation y. $^*$ For the $k=e\cos \varpi$ and $h=e\sin\varpi$ variables, an additional prior was added to enforce a uniform distribution for their module $e$. } 
\end{table*}

 \begin{figure}[!ht]
\begin{center}
\includegraphics[width=0.49\textwidth]{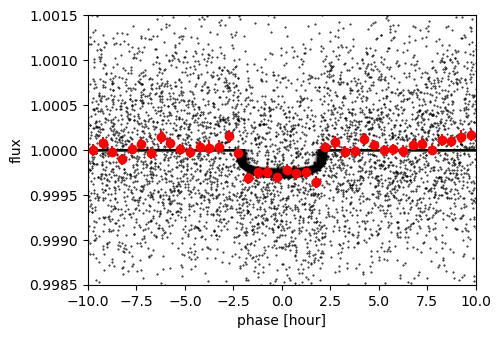}
\includegraphics[width=0.49\textwidth]{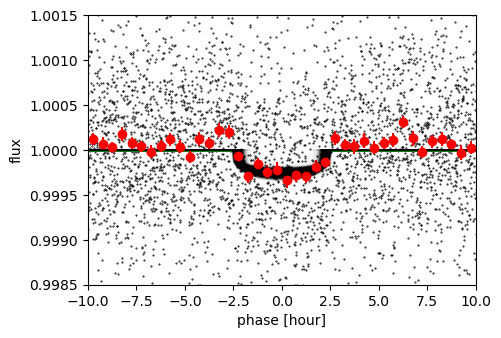}
\caption{\label{fig:KOI4772stacked} Stacked transits of Kepler-1705b (\textit{top}) and Kepler-1705c (\textit{bottom}). TTV-folded data are shown as black dots, bins of 30 mins  are shown in red, and 1000 randomly selected posterior samples in black lines.}
\end{center}
\end{figure}

\begin{table} 
\caption{Transit timings of Kepler-1705b and c in BJD-2454833.0. } 
\label{tab:transittimings} 
\centering 
\begin{tabular}{lll} 
median & $-\sigma$ & $+\sigma$\\ 
\hline\hline 
\multicolumn{3}{c}{Kepler-1705b}\\ 
\hline 
365.2143&-0.0109&0.0093\\ 
374.2501&-0.0103&0.0091\\ 
383.2838&-0.0099&0.0086\\ 
\multicolumn{3}{c}{ ... }\\ 
\hline\hline 
\multicolumn{3}{c}{Kepler-1705c}\\ 
\hline 
374.3956&-0.0176&0.0201\\ 
385.6777&-0.0172&0.0194\\ 
396.9589&-0.0169&0.0185\\ 
\multicolumn{3}{c}{ ... }\\ 
\end{tabular} 
\tablefoot{$-\sigma$ corresponds to  the 0.15865 quantile minus the median value, $+\sigma$ corresponds to the 0.84135 quantile minus the median value. The full table is available online for download.}  
\end{table}

The fit of the data was performed in two steps: a preliminary fit of the transit timings to the timing proxy shown in Fig. \ref{fig:KOI4772_TTV}, followed by a photodynamic analysis of the light curve. Both steps use the {\ttfamily TTVfast} algorithm \citep{DeAgHo2014} for the computation of the transit timing for a set of initial conditions, and the samsam\footnote{\url{https://gitlab.unige.ch/Jean-Baptiste.Delisle/samsam}} MCMC algorithm \citep[see][]{Delisle2018} to sample the posteriors. The light curve model was obtained by modelling the transit of each planet using the {\ttfamily batman} package \citep{batman}, shifting its centre to the dates of transits computed with {\ttfamily TTVfast}. The supersampling parameter was set to $29.42$ minutes to account for the long exposure of the dataset. Remaining long-term trends were modelled by a Gaussian process with a Matérn 3/2 kernel whose timescale was forced to be above one day to avoid interfering with the modelled transits. A jitter term was also added to all photometric measurements. The stellar parameters were retrieved from the Gaia-Kepler Stellar Properties Catalog \citep{Berger2020} and given in Table \ref{tab:stellarParam}. The effective temperature, ${\log g,}$ and metallicity were used to compute the quadratic limb-darkening coefficients $u_1$ and $u_2$ adapted to the Kepler spacecraft using {\ttfamily LDCU}\footnote{https://github.com/delinea/LDCU}. The fit was performed on the flattened DPCSAP flux, removing the data points at less than twice the transit duration of each transit of the 3.38-day candidate.

The resulting fitted and derived posteriors are given in Table \ref{tab:planetParam}. The fit converged toward a pair of similar-sized planets with masses of $4.61_{-0.45}^{+0.54}$ and $5.60_{-0.58}^{+0.66}$M$_{Earth}$, and radii of $2.04_{-0.14}^{+0.12}$ and $2.05_{-0.15}^{+0.14}$R$_{Earth}$, respectively. The eccentricities and longitude of periastron are strongly degenerate, as discussed in Sect. \ref{sec:TTVdeg}. The TTV-corrected stacked light curve of each planet can be seen in Fig. \ref{fig:KOI4772stacked}. The TTVs corresponding to 300 randomly chosen samples of the photodynamic fit are shown in Fig. \ref{fig:KOI4772_TTV} for comparison to the proxy generated from the confidence matrix.

The two planets are detected with S/Ns above 8.5. In addition, the anti-phased nature of the TTV signals shown in Fig. \ref{fig:KOI4772_TTV}, the fact that these timings can be reproduced by a model of two planets in resonance, and the masses determined by the TTV analysis that are consistent with the derived radii, are independent confirmations of the planetary nature of the signals.

\section{The resonant pair of Kepler-1705}

\label{sec:dyn}

 \begin{figure}[!ht]
\begin{center}
\includegraphics[width=0.49\textwidth]{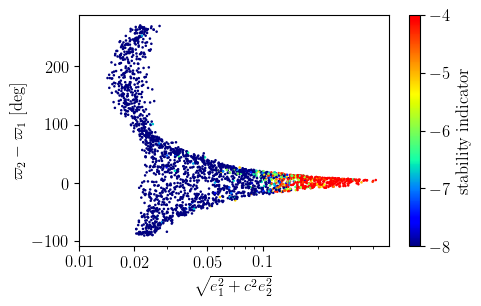}
\includegraphics[width=0.49\textwidth]{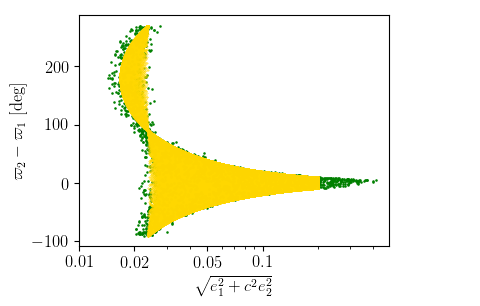}
\includegraphics[width=0.49\textwidth]{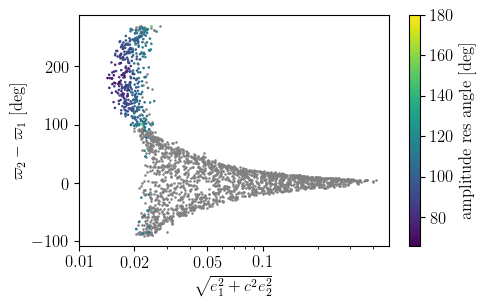}
\caption{\label{fig:KOI4772stab} \textit{Top:} Stability of the trajectories integrated from 2300 posterior samples, ranging from stable (blue) to unstable (red). \textit{Middle:} Green dots show the projection of 2300 randomly selected samples of the posterior summarised in Table \ref{tab:planetParam}. Gold dots show the theoretical posterior discussed in section \ref{sec:TTVdeg}.  \textit{Bottom:} Smallest semi-amplitude between the two resonant angles (eq. \ref{resangs}). Grey dots indicate a circulation of both of them.}
\end{center}
\end{figure}

In this section, we study a subpopulation of 2300 randomly chosen samples from the posteriors presented in Table \ref{tab:planetParam}. Figure~\ref{fig:KOI4772stab} shows the projection of these samples in the $(\sqrt{e_1^2+c^2e_2^2},\Delta \varpi = \varpi_2-\varpi_1)$ plane ($c$ is defined in section \ref{sec:TTVdeg}). The $e_j$ and $\varpi_j$ represented here are the fitted initial conditions at the date 2455196.2294 BJD.

\subsection{Stability}  
We verified the stability of the posteriors using the frequency analysis criterion \citep{La90,La93}, using the same implementation as in \cite{Leleu2021}. For this stability analysis, KOI4772.01 was added to the system using the ephemerides from the NASA exoplanet archive\footnote{https://exoplanetarchive.ipac.caltech.edu/cgi-bin/TblView/nph-tblView?app=ExoTbls\&config=cumulative}. Assuming an Earth-like density, we used a mass of $1.2e\!-\!05M_\star$. The top panel of Fig. \ref{fig:KOI4772stab} shows the resulting criterion for each initial condition, for integration over $10^5$ years. We find that the low-eccentricity part of the posterior is stable for more than $10^6$ years, which together correspond to more than 35 billion orbits of the resonant pair. However, a significant part of the high-eccentricity posterior is unstable on a short timescale (red dots in the top panel of Fig. \ref{fig:KOI4772stab}). This instability is due to the presence of the inner planet; running the same analysis without KOI4772.01 gives a fully stable posterior. 

%\subsection{Resonant state}  

\subsection{Dynamics and TTV signal degeneracy}
\label{sec:TTVdeg}

The middle panel of Figure~\ref{fig:KOI4772stab} shows the 2300 randomly chosen samples from the posterior presented in Table \ref{tab:planetParam}. Plotted is a projection of these samples in the $(\sqrt{e_1^2+c^2e_2^2},\Delta\varpi=\varpi_2-\varpi_1)$ plane
(green dots), where $_1$ and $_2$ refer to the inner and outer planet of the pair, respectively, and $c$ is defined below,
together with the distribution expected on theoretical grounds (gold dots). The latter can be understood as follows.

\begin{figure}
\centering
\includegraphics[width=80mm]{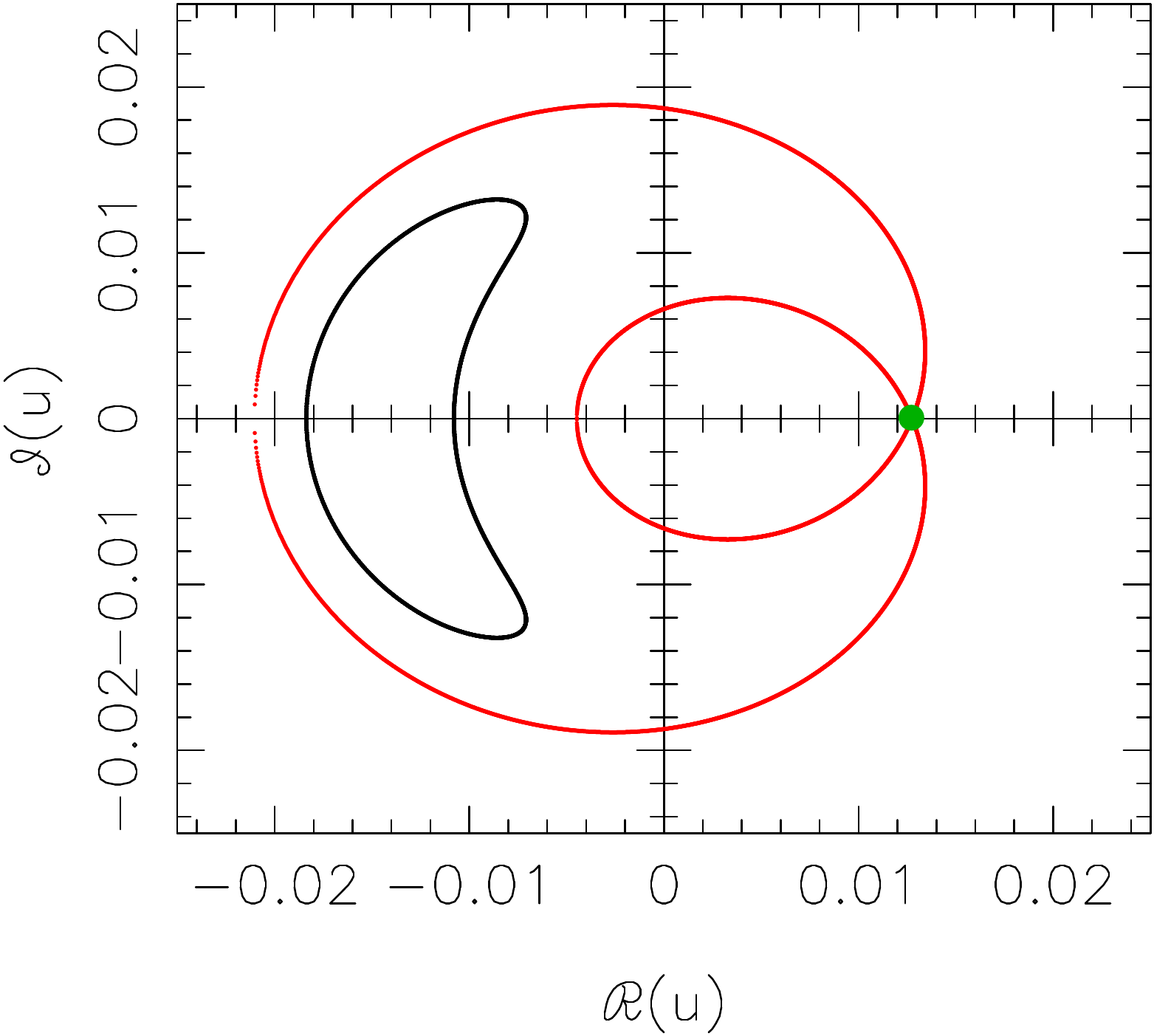}
\caption{A phase-space plot showing the variation of the real and imaginary parts of the transformation variable $u$ (black curve: see eq.~(\ref{eq:uv}) for its definition in terms of the eccentricities and resonance angles).
The best-fit solution has been used for the initial conditions of the black curve. As $u$ is constant over its theoretical distribution (yellow points in the second panel of Fig.~\ref{fig:KOI4772stab}), and varies little over its MCMC distribution (green points), the black curve represents in practice all points of the posterior.
The phase space in which a system resides is resonant when there exists a solution containing a hyperbolic fixed point (green point in this figure) which has the
same angular momentum as the system but a different energy. The {\it separatrix} (red curve) separates {\it librating} solutions from 
{\it circulating} solutions. The Kepler-1705 system is formally resonant because the phase space is resonant;
it is in the librating state because it resides inside the separatrix.
Being in the librating state allows determination of the planetary masses, independent
of the eccentricities and the resonant state of the phase space (see text for discussion).
We note that these curves are not surface-of-section projections but are true one-dimensional curves.
}
\label{fig:KOI4772phase}
\end{figure}

In addition to the usual energy and angular momentum integrals,
coplanar systems near first-order commensurabilities have an additional approximate integral
which is exact to first-order in the eccentricities when  only the two resonant harmonics are
included in the disturbing function \citep{sessin1984,henrard1986,wisdom1986}. Noting the resonant angles for the 5:4 resonance:
\be
\phi_1= 4 \lambda_1-5 \lambda_2 + \varpi_1 \ ; \ 
\phi_2= 4 \lambda_1-5 \lambda_2 + \varpi_2\, ,
\label{resangs}
\ee
the integral is revealed by transforming the complex variables 
$z_1=e_1\,{\rm e}^{i\phi_1}$ and $z_2=e_2\,{\rm e}^{i\phi_2}$ to
\be
u=z_1+c z_2
\hspace{0.5cm}{\rm and}\hspace{0.5cm}
v=cz_1-gz_2,
\label{eq:uv}
\ee
where $c$ is a function of the Laplace coefficients and has a value of $c=-1.119$ for the 5:4 commensurability, and $g=(m_2/m_1)\sqrt{a_2/a_1}$
(Mardling, in prep). We note that different scalings are used by different authors; for the purpose of this paper
we use a scaling that reflects the values of the eccentricities. 
In terms of the transformed variables, 
the additional integral is simply $vv^*={\rm const}$, and its existence results in a one-dimensional curve
in the $(\Re(u),\Im(u))$ plane, where $\Re(u)$ and $\Im(u)$ are the real and imaginary parts of $u$ 
(black curve in Figure~\ref{fig:KOI4772phase}). 
One can show that the TTV signal contains information about $u$ only, that is, it
is blind to $v$ at first order in eccentricity (Mardling, in prep). As $u$ involves a linear combination of
the quantities $e_1{\rm e}^{i\varpi_1}$ and $e_2{\rm e}^{i\varpi_2}$, solving the TTV-inverse problem
results in degeneracies in the eccentricities and the reference-frame-independent angle
$\phi_2-\phi_1=\varpi_2-\varpi_1$. 
Thus, while $u$ varies little over the posterior of a solution, $v$ is completely free, depending only
on the imposed priors for the components of $e_1{\rm e}^{i\varpi_1}$ and $e_2{\rm e}^{i\varpi_2}$. The yellow dots in the middle panel of Figure~\ref{fig:KOI4772stab}
were generated by taking the best-fit solution, holding $u$ constant, and varying
the real and imaginary parts of $v$ separately over the range $[-0.2,0.2]$.
The resulting distribution closely matches the posterior for the solution.

We highlight the fact that the distribution is composed of two distinct components: points with $-90^o<\Delta\varpi<90^o$
and points with $90^o<\Delta\varpi<270^o$,
corresponding to systems for which $\Delta\varpi$ librates around zero and $180^o$ respectively, but with circulation occurring near the boundary between the two.

As the stationary (fixed-point) values of $\phi_1$ and $\phi_2$ are zero and $180^o$ respectively,
libration of both angles simultaneously
can only occur around these values, and as a result $\Delta\varpi$ librates around $180^o$ in this circumstance. On the other hand, libration of $\Delta\varpi$ around zero can only occur when both angles {\it circulate}
and have similar values.

Figure~\ref{fig:KOI4772phase} shows that the  Kepler-1705 system is in the {\it librating resonant} state, determined by the behaviour of the `resonant argument', $\psi,$ defined such that $u=|u|{\rm e}^{i\psi}$.
Librating systems (whether they are resonant or not) are characterised by {two} distinct non-commensurate resonant frequencies, one commonly referred
to as the {super-frequency} \citep{Lithwick2012}, and a second which is usually significantly higher
(Mardling, in prep; see also Nesvorny \& Vokrouhlicky 2016). We note that these two frequencies are nearly degenerate
for {circulating} systems. Therefore, if the observing baseline of a librating system is sufficiently long for both frequencies to be detected, there should be enough information in the signal to determine the 
planet masses {independently of the eccentricities}, that is, the
mass--eccentricity degeneracy which plagues many TTV systems is broken (Mardling, in prep).  We note that this does not depend on detecting the zeroth-order-in-eccentricity `chopping' component of the TTV signal, which, although depending on the planet mass only, is generally much smaller than the resonant component (see, e.g. Linial et al 2018).
For the  Kepler-1705 system, the corresponding super-period is around 4000 d, while the second resonant
period is around 1000 d; it is the latter which dominates the TTV signal as Figure~\ref{fig:KOI4772_TTV} shows.
Although the super-period is around four times the observing baseline, its effect on the signal
allows for clear determination of the planet masses within 10\%.

We briefly compare these results with the Hamiltonian formulation of the second fundamental model for resonance of \citet{HenLe1983}. Using the one-degree-of-freedom model of first-order resonances presented in \cite{DePaHo2013}, we computed the value of the Hamiltonian parameter $\Gamma'$ (eq. 36 in the aforementioned paper), and found it to be relatively well constrained over the whole posterior, $\Gamma'=2.43_{-0.35}^{+0.37}$, despite the degeneracies illustrated in Fig. \ref{fig:KOI4772stab}. We computed the position of the separatrix for all the dots displayed in this figure. In all cases, the trajectories lie within the separatrix, ensuring that the system lies inside the 5:4 MMR (see Figure~\ref{fig:KOI4772phase}).

Finally, we note that a system can be in the librating resonant state even when both resonant angles circulate;
this is the case when $\Delta\varpi$ librates around zero. 
Thus, the standard conclusion drawn that a system is {not} resonant unless at least one resonant angle librates
is erroneous in this situation.

Classifying Kepler (near-)resonant pairs with respect to the period of the inner planet in the pair, \cite{delisle_2014_tidal} showed that systems where the period of the inner planet of the pair is smaller than 15 days tend to be near-resonant rather than inside the resonance, which can be explained by the tidal interaction with the star. The resonant state of the Kepler-1705 system can be explained by a lower dissipation in the planets, but could also be linked to the age of the system ($t_{\star}=2.59_{-1.61}^{+1.94}$ Gyr), if for example this latter is not yet sufficient to allow the tidal evolution to be fully effective.

\subsection{Planetary spin dynamics}  
\label{sec:spin}

The spin of close-in, nearly circular planets is often assumed to be synchronised with the orbital motion
due to strong tidal dissipation in the planets.
In the case of Kepler-1705, the outer planet has a period of 11.28~d,
and so one would indeed expect all the planetary spins to be in a synchronous state.
However, planet--planet perturbations can drive the spin of rocky planets
into asynchronous or chaotic states, even for nearly circular orbits \citep[see][]{CoRo2013,LeRoCo2016,DeCoLeRo2017,correia_2019_spinorbit}.
In particular, as shown by \citet{DeCoLeRo2017},
the TTVs are a very good probe with which to study the spin dynamics,
because both TTVs and spin--orbit dynamics are dominated by the perturbations
on the mean longitude of the planets.
Planets with large TTVs, such as  Kepler-1705b and c, thus also undergo
strong perturbations in their spin dynamics,
and could therefore be captured in non-synchronous states or could exhibit a chaotic evolution.
At first approximation, the planet--planet perturbations introduce
two new spin--orbit resonances surrounding the synchronous resonance \citep{CoRo2013}.

We studied the impact of these planet--planet perturbations
on the spin evolution of Kepler-1705b and c.
The details of this analysis are presented in Appendix~\ref{sec:app_spin},
and we briefly summarise our findings here.
For  Kepler-1705b, a permanent capture in the synchronous resonance is the most probable scenario, while a capture in the non-synchronous resonances does not seem possible.
However, the spin of  Kepler-1705b could undergo a chaotic evolution for a long time before
tidal dissipation finally brings it to permanent capture in the synchronous resonance
\citep[see][]{DeCoLeRo2017}. For  Kepler-1705c, a permanent capture in any of the three resonances (synchronous, super-synchronous, or sub-synchronous) is possible.
As for  Kepler-1705b, the spin of  Kepler-1705c could also undergo a chaotic evolution for a long time before
being permanently captured in one of these resonances.

\section{Discussion and conclusion}
\label{sec:discncon}
\subsection{Choices and caveats}
\label{sec:caveats}

In this paper we present a proof of concept: deep learning can be used to identify the track of a planet in a river diagram regardless of the presence of TTVs. To present this method, numerous choices were made, notably regarding the architecture of the neural network. Parts of the experimentation that was conducted prior to these choices are detailed in Appendix \ref{sec:experiments}. The cost function for the training of the model is also arbitrary, and is generally a trade-off between the recall (not missing the signature of a planet) and the false positive rate, both for the semantic segmentation (pixel of the river diagram belonging to a transit), and for the classification (track of the planet in the input matrix). Typically, cleaner periodograms can be obtained by reducing the false positive rate of the classification. On the other hand, a higher recall on the classification helps to recover planets with a lower S/N. The choice of training set can also influence the sensitivity of the model. In this paper we chose to train the model on a wide variety of systems in, or close to various MMRs, around any kind of star. Training per stellar type, or for a given orbital configuration, will probably provide better results for the dedicated task, provided that a large enough training
set can be constructed.

Both the river diagram classifier and the pixel-level vetting of the RIVERS.deep matrix return confidence of the model on their task. These confidences are not likelihoods or probabilities. In that sense, this method can only be used to detect promising signals in large datasets, which in turn need to be analysed with a `classical' study of the light curve, as performed in section \ref{sec:fit}.   
%benefit from the statistical background of tools such as the BLS or QATS algorithm that are based on $\chi_2$ minimisation. 

%Neural network decisions are not easily interpretable. In that sense, the classification 

\subsection{Summary and Conclusion}
\label{sec:conclusion}

For planets that are too small to induce individually detectable transits, TTVs could lead to erroneous estimations of the transit depth and duration, or even the absence of detection. For such planets, we present a method, using deep learning with tasks of semantic segmentation and image classification, to recover small planets with transit surveys with an approach that is robust to TTVs. This approach does not rely on the recognition of individual transits, nor on the stacking of the light curve to identify the candidate, but rather on the track that is drawn by the planet in a river diagram. We show that, in addition to the orbital period, individual transit timings can be estimated by the semantic segmentation task. 

We illustrate the method by the detection and confirmation of a pair of super-Earths in a 5:4 MMR around Kepler-1705. Each of these planets has TTVs of $\sim 10$ h amplitude, and individual transit S/N of $\sim 1$. These planets do not appear when using the BLS algorithm. When comparing the resonant pair of Kepler-1705 to the previously known planets with large TTVs in Fig. \ref{fig:kepler}, it appears that the RIVERS.deep method presented in this paper is indeed able to recover planets that had three times lower individual transit S/N  than any other known planet with TTVs of more than 3 hours. The method therefore has the potential to alleviate the TTV bias in transit surveys. As this bias is due to orbital perturbations that happen on timescales longer than the orbital period, it is especially useful for observations with long baselines, such as Kepler, the polar observations of TESS, and the upcoming PLATO mission.

We find that Kepler-1705 is in a librating resonant state, and that this state varies little over the entire posterior of the photodynamic fit. Moreover, because the system is in this state the planet masses are well constrained, in contrast to systems in a circulating state for which the planet masses and eccentricities are degenerate. On the other hand, the eccentricities, resonant angles, and longitudes of periastron are not constrained, except for a lower limit for the eccentricities of around 0.01. This results from the fact that TTVs are sensitive to a component of the combined complex eccentricities only.

We also show that the significant orbital perturbations induced on Kepler-1705b and Kepler-1705c can significantly impact their spin dynamics. In both cases, a large chaotic area surrounds the synchronous spin--orbit resonance. The spin of the inner planet of the resonant pair should settle into the synchronous state after crossing the chaotic area. The outer planet can be locked either in the synchronous spin--orbit resonance, or in one of the sub- or super-synchronous resonances which originate from the orbital perturbation. This would have important consequences for the climate of the planet because a non-synchronous spin implies that the flux of the star is spread over the whole planetary surface. Regardless of which spin--orbit resonance these planets are trapped in, the orbital resonant motion induces forced oscillations of the spin with respect to the planet--star direction, which is a source of tidal dissipation.

The paucity of planets in MMR in the Kepler dataset is a subject that has been extensively studied in the last decade \citep[e.g.][]{DeLaCoBo2012,Fabrycky2014,GoSc2014}. Our study is a first step towards showing that part of this trend might be due to an observational bias. Recovering these small, (near-)resonant planets is valuable for our understanding of the formation and evolution of planetary systems, because the formation and evolution of resonant configurations are the markers of the dissipative evolution of planetary systems \citep[see e.g.][]{HenLe1983,LePe2002,TePa2007,PaTe2010,DeLaCoBo2012,CoDeLa2018}. In addition, for faint stars such as Kepler-1705, TTVs are currently our only means to estimate the mass of the planets, and hence their density and internal structures.

\begin{acknowledgements}
This work has been carried out within the framework of the National Centre of Competence in Research PlanetS supported by the Swiss National Science Foundation and benefited from the seed-funding program of the Technology Platform of PlanetS. The authors acknowledge the financial support of the SNSF.
\end{acknowledgements}

\bibliographystyle{aa}
\bibliography{biblio_RIVERS_deep}

\appendix

\section{TTV timescales}
\label{sec:TTV}

A two-body MMR is an orbital configuration in which the period of a pair of planets satisfies the following relation: $P_2/P_1\simeq (k+q)/k$, where $k$ and $q$ are integers, the latter being the order of the resonance. For a pair of planets inside a first-order ($q=1$) resonance, TTVs happen on a timescale:
\begin{equation}
P_\text{TTV, $q=1$} \propto f (m_p / m_\star)^{-2/3} P_{orb} \, ,
\label{eq:TTVres}
\end{equation}
where $P_{orb}$ is comparable to the orbital period of the planets, $m_p$ is comparable to the mass of the planets, and $f$ is a factor or order 1 which increases as the libration amplitude increases \citep[][]{NeVo2016}. In the co-orbital resonance ($q=0$), TTVs happen on the timescale \citep{VoNe2014,Leleu2019}: 
\begin{equation}
P_\text{TTV, $q=0$} \propto f (m_p / m_\star)^{-1/2} P_{orb} \, .
\label{eq:TTVres}
\end{equation}

For a pair of planets near but not in a MMR of order $1$ or $2$, the TTVs happen on the period associated with the distance to the exact resonance $P_2/P_1=(k+q)/k$ in the frequency space:
\begin{equation}
P_\text{TTV, near res} = \frac{1}{(k+q)/P_2 -k/P_1} \, .
\label{eq:TTVineq}
\end{equation}
In the neighbourhood of first-order MMRs, these TTVs are sinusoidal at first order in eccentricities \citep{Lithwick2012,AgolDeck2016,Mardling2018}. A particular case of Eq.~\ref{eq:TTVineq} for $q=0$ is referred to as the `chopping term', which happens mainly for close pairs of planets near conjunction \citep{NeVo2014}.

Then, on longer timescales the evolution of eccentricities and inclinations ---and associated angles--- produce TTVs  on secular timescales \citep[see the Laplace-Lagrange theory][]{MuDe1999}:
\begin{equation}
P_\text{TTV, sec} \propto (m_p / m_\star)^{-1} P_{orb} \, .
\label{eq:TTVsec}
\end{equation}
Three-body resonances can also produce variation over this timescale. The effect of these configurations can add up linearly at first order. 

\section{Effect of TTVs on transit depth}
\label{ap:smearing}

\begin{figure}[h!]
    \centering
    \includegraphics[width=0.49\textwidth]{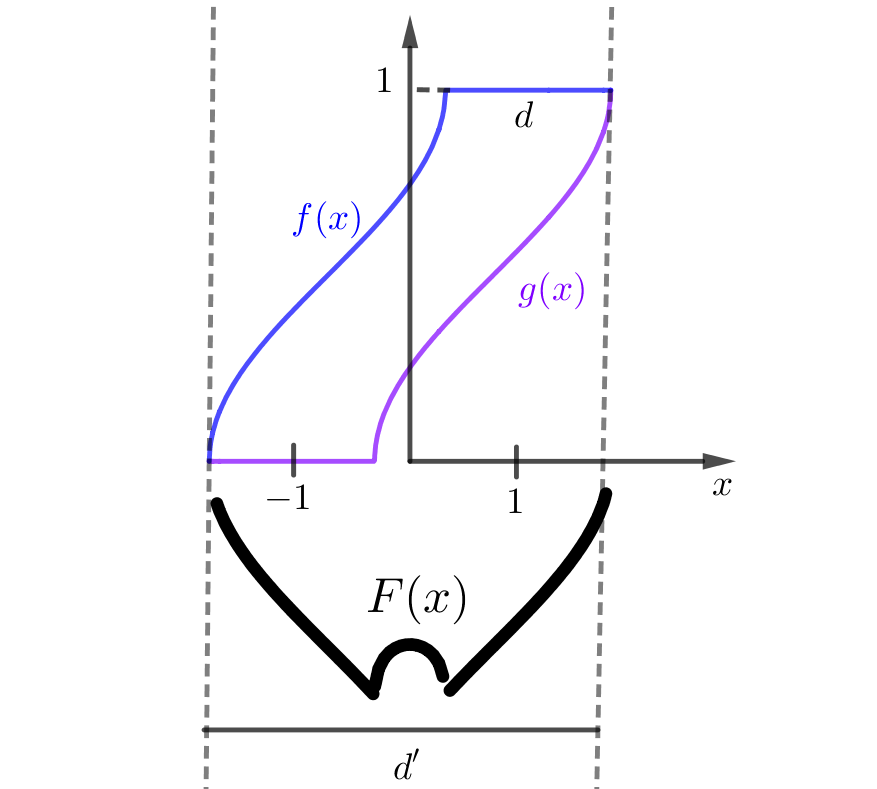}
    \caption{\label{fig:DetphSmear} $f(x)$ gives the fraction of the time where $x$ is after the ingress, $g(x)$ gives the fraction of the time where $x$ is after the egress. $F(x)=g(x)-f(x)$ shows the shape of the stacked transit.}
\end{figure}

We estimate here the effect of TTVs on the estimated transit depth, when the TTVs are not modelled. We assume sinusoidal TTVs of period $2$ and of semi-amplitude $\sigma_\text{TTV}/2$. We note the duration of the underlying planetary transit $T_\text{transit}$. We normalise the phase of transit by the TTV semi-amplitude $\sigma_\text{TTV}/2$; and we parametrise the normalised phase by $x$, taking $x=0$ at the average mid-transit position (see Fig. \ref{fig:DetphSmear}). We assume the actual transits to be box-shaped, of depth $D,$ and of normalised duration $d=T_\text{transit}/(\sigma_\text{TTV}/2)$. Due to the symmetry of the $\sin$
%(\theta+\pi/2)$ function with respect to $\theta=0$
function, we compute the average effect over half a TTV period. % $\theta \in [0,\pi]$.
Noting $f(x)$ the fraction of the time when $x$ is after the ingress, and $g(x)$ the fraction of the time when $x$ is after the egress, we have (see Fig. \ref{fig:DetphSmear}):
\be
\begin{aligned}
f(x) =& \arcsin (x +d/2)/\pi + 1/2 & \text{for}\ x\leq 1 -d/2 \\
f(x) =& 1 & \text{for}\ x\geq 1 -d/2
\end{aligned}
\ee
and
\be
\begin{aligned}
g(x) =& 0 & \text{for}\ x\leq -1 +d/2 \\
g(x) =& \arcsin (x - d/2) /\pi+ 1/2  & \text{for}\ x\geq -1 + d/2.
\end{aligned}
\ee
The flux loss due to the stacked transit at phase $x$ is given by $DF(x)$, where
\be
\begin{aligned}
F(x) = g(x)-f(x)
\end{aligned}
,\ee
which is the opposite of the fraction of time the phase $x$ is in transit, see Fig. \ref{fig:DetphSmear}. The recovered transit depth depends on the chosen stacked transit duration $d'$:
\be
\begin{aligned}
\text{Smeared Depth} &= \frac{D}{d'}\int^{d'/2}_{-d'/2} -F(x) dx 
\end{aligned}
\label{eq:smearedfull}
.\ee
Choosing $d'=2+d$, which results in a stacked transit that contains all the phases affected by the underlying transits as shown in Fig. \ref{fig:DetphSmear}, eq. \ref{eq:smearedfull} simplifies to:
\be
\begin{aligned}
\text{Smeared Depth} &= dD/(2+d)\, .
\end{aligned}
\ee

We note that here we assume that the transits are stacked following the `true' mean period of the planet. If the observation`s cover only a fraction of the TTV period, a fit of the data can identify an instantaneous period that will lessen the smearing. %In addition, we assume here that the considered transit duration span $-1-d/2 \leq x \leq 1+d/2$, but modeling a shorter transit can reduce the effect of the smearing on the depth.  

\section{Model description}
\label{ap:model}

\subsection{Model architecture}

\subsubsection{Semantic segmentation model}

\label{subsubsec:semseg}

\begin{figure}
    \centering
    \includegraphics[width=0.24\textwidth]{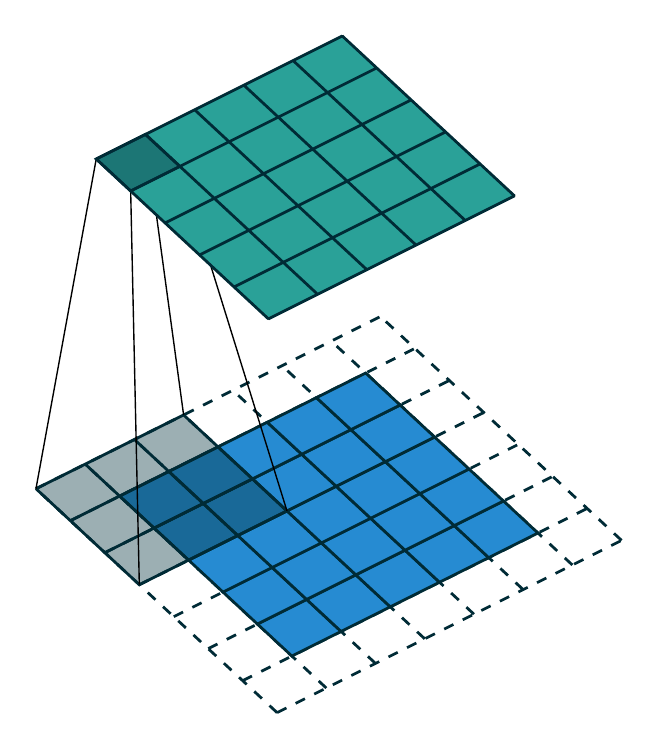}
    \includegraphics[width=0.24\textwidth]{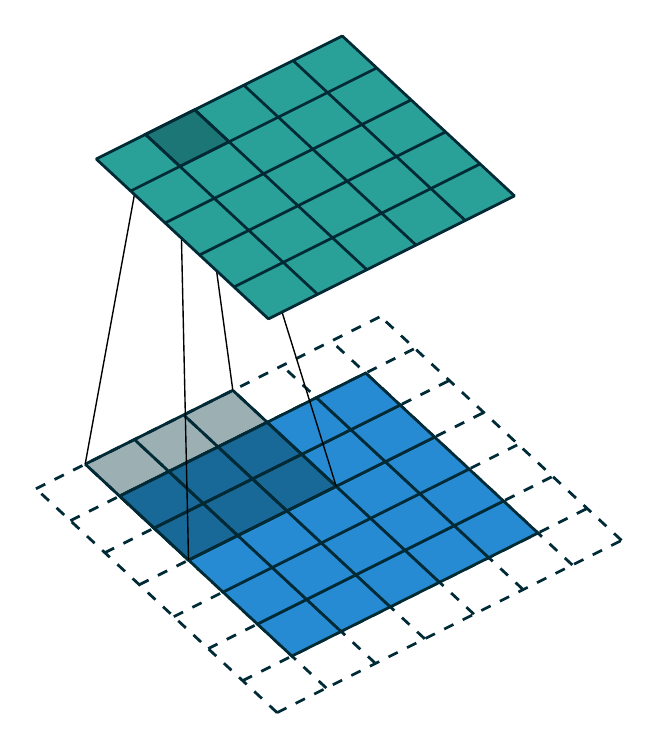}
    \caption{\label{fig:conv_3x3} Application of a $3 \times 3$
      convolution filter (gray) on a $5 \times 5$ input (blue) to
      generate a $5 \times 5$ output (teal). To be able to center the
      filter on every pixel of the input (and therefore keep the same
      dimensions), we use one pixel of padding (dashed) with value 0
      around the image. }
\end{figure}

The main building block of the semantic segmentation model is the
\textit{convolution layer} \citep{dumoulin2016guide}. A convolution
layer is composed of multiple \textit{convolution filters} which are
matrices of floating point values (in our case of shape $3 \times
3$). Applying a filter to an image (as depicted in
Fig. \ref{fig:conv_3x3}) consists in sliding it over each pixel of the
input and computing a weighted sum of the overlapped region. The
weight associated to each pixel of the input is given by the
corresponding value in the filter. During the training of the neural
network, each filter will learn to highlight \textit{local patterns}
that will be useful for the next neural layers to work with. The goal
of \textit{deep} learning approaches is to let the model discover a
\textit{deep hierarchy} of features that enables it to solve the task
at hand.

\begin{figure}
  \begin{center}
    \includegraphics[width = 7cm]{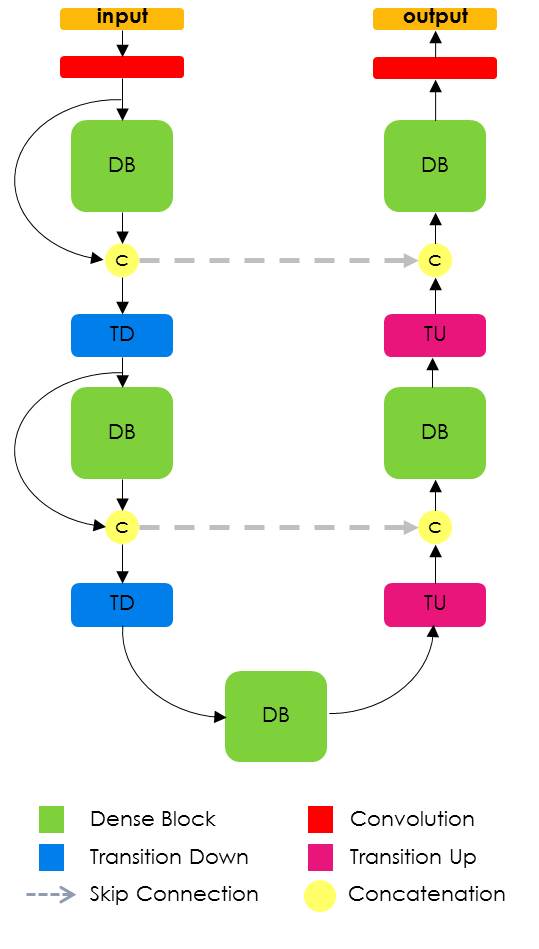}
  \end{center}
  \caption{\label{fig:tiramisu} Architecture of the Tiramisu neural
    network used for the semantic segmentation task. The
    \textit{encoder} (left part of the figure) is a subnetwork that
    works with an increasingly coarse representation of the data by
    using transition down blocks that reduce the resolution. The
    \textit{decoder} (right part of the model) is a subnetwork that
    progressively combines low-resolution information coming from
    transition up blocks and high-resolution information coming from
    the encoder by skip connections. The \textit{bottleneck} (dense block at 
    the bottom of the figure) is used to refine the high level of 
    understanding that the model has of the input.}
\end{figure}

The model used for the semantic segmentation task is a fully
convolutional DenseNet \citep{jegou2017one} illustrated in
Fig.~\ref{fig:tiramisu}.

This architecture is particularly well suited to this job as there are
multiple `paths' from input to output in the model. Short paths are
useful to analyse \textit{small local neighbourhoods} and predict on a
pixel-by-pixel basis which ones are more likely to belong to a
transit. Long paths are more focused on deciding whether the
\textit{global signal structure} is coherent. By merging these two
sources of features, the model is able to produce high-quality transit
masks by looking for areas with lower average flux and a global
structure that could be explained by a planet.

\begin{figure}
  \begin{center}
    \includegraphics[width = 4cm]{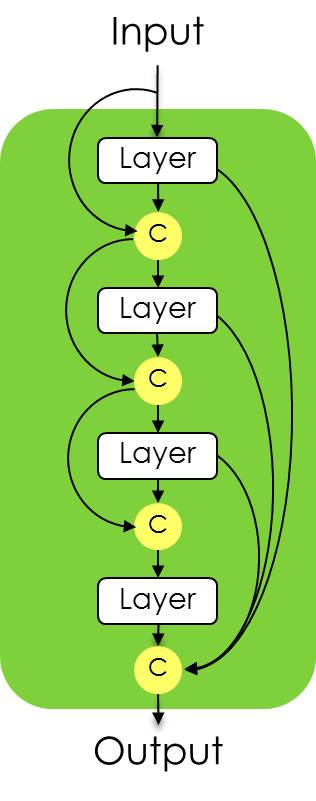}
  \end{center}
  \caption{\label{fig:dense_block} Architecture of one of the dense blocks that
    compose the Tiramisu model.}
\end{figure}

The first convolution layer of the model includes eight filters and the filter
growth rate is of five (each convolution layer has five more filters than
the previous one). The encoder and the decoder are both composed of five
\textit{dense blocks}. Dense blocks \citep{huang2017densely},
illustrated in Fig.~\ref{fig:dense_block}, are sequences of
convolution layers with skip connections allowing the output to
contain information from multiple levels of the feature hierarchy
discovered by the network. 

In the encoder, the dense blocks are interlaced with transition down blocks. The objective of these blocks is to reduce the resolution of the signal being processed in order to increase the scale of the pattern the next layers will process. This is performed using a maximum pooling operation \citep{scherer2010evaluation}. This layer divides the width and the height of the signal by a factor of two by tiling its input using $2 \times 2$ cells and replacing each cell by the maximum value it contains.

The bottleneck layer at the transition between the encoder and the decoder is used to refine the large-scale understanding the
model has of the input, and is composed of a single dense block with eight 
convolution layers.

In the decoder, the dense blocks are interlaced with transition up blocks. The objective of these blocks is to increase the resolution of the signal being processed in order for the model to be able to produce an output of the same shape as its input (therefore producing a prediction for each of the pixels). This is performed using a variant of the convolution layer called a \textit{transposed convolution} or \textit{fractionally strided convolution} \citep{zeiler2010deconvolutional, dumoulin2016guide}. As its name suggests, this operation increases the resolution of its input by applying a convolution operation in which the filter is moved by a fraction of a step. In our model, the filter is moved by half a step for each computation, therefore doubling the width and height of the signal.

\subsubsection{Classifier model}

The classifier model takes the output of the semantic segmentation
model as input and predicts whether or not the river diagram contains
a perturbation.

The classifier is a convolutional neural network. Its base components
are convolution blocks and linear blocks. A \textit{convolution block}
is composed of the following pieces:
\begin{itemize}
\item a convolution layer as defined in Section~\ref{subsubsec:semseg},
\item a 20\% dropout layer \citep{JMLR:v15:srivastava14a} which
  randomly replaces 20\% of the values of its input by 0 in order to
  improve generalisation and reduce dataset memorisation by the model (overfitting),
\item a batch normalisation layer \citep{ioffe2015batch} that
  accelerates the training procedure of the model by normalising its
  input,
\item a ReLU activation function \citep{glorot2011deep} which is the
  simple non-linear function $f(x) = \text{max}(x, 0)$. This function
  allows the model to approximate non-linear functions as it is
  otherwise entirely composed of linear combinations.
\end{itemize}

\noindent A \textit{linear block} is composed of the following pieces:
\begin{itemize}
\item a linear (or fully connected) layer
\item a 20\% dropout layer,
\item a batch normalisation layer,
\item a ReLU activation, except for the very last block which uses a
  softmax activation. The softmax function normalises the raw output of the last layer into a probability distribution over all the classes of the classification task.
\end{itemize}

The sequence of convolution blocks is interlaced with maximum pooling operations \citep{scherer2010evaluation}. This operation reduces the height and width of the array it is processing by a factor of two. It is used for two main reasons: (1) It allows us to mitigate the computation cost of increasing the number of filters in the following convolution blocks. We increase the number of filters in order to be able to recognise a greater variety of patterns. (2) It improves the translation invariance of the model. We would like the model decisions to be the same regardless of the positions of the planet track in the diagram. The maximum pooling operation is designed to loosen the spatial position of the patterns.

The sequence of blocks in the feature-extraction part of the model is given in Table \ref{tab:seqblocks}.

\begin{table} 
\caption{Sequence of blocks in the feature-extraction part of the model} 
\label{tab:seqblocks} 
\centering 
  \begin{tabular}{| c| c| c |}
    \hline
    layer & input channels & output channels \\
    \hline
    \hline
    conv\_block\_1 & 2 & 32 \\
    conv\_block\_2 & 32 & 32 \\
    conv\_block\_3 & 32 & 32 \\
    max\_pool\_2d & &\\
    \hline
    conv\_block\_4 & 32 & 64 \\
    conv\_block\_5 & 64 & 64 \\
    conv\_block\_6 & 64 & 64 \\
    max\_pool\_2d & &\\
    \hline
    conv\_block\_7 & 64 & 128 \\
    conv\_block\_8 & 128 & 128 \\
    conv\_block\_9 & 128 & 128 \\
    max\_pool\_2d & & \\
    \hline
    conv\_block\_10 & 128 & 256 \\
    conv\_block\_11 & 256 & 256 \\
    conv\_block\_12 & 256 & 256 \\
    max\_pool\_2d & &\\
    \hline
  \end{tabular}
\end{table} 

After flattening the output of these layers into a 1D vector, a
sequence of three linear blocks with 256, 256, and 2 nodes is applied.

\subsection{Training loss}

When training neural network models, we have to define a differentiable
\textit{loss function} that quantifies how good the model prediction
is compared to the desired value from the training dataset, the
\textit{label} or \textit{target}. The loss function used to train our
model is dependent on two sublosses, one for the semantic segmentation
(pixel class prediction) and one for the classification (diagram class
prediction).

The \textit{binary cross entropy} loss ($BCE$) is defined as follows:

\begin{equation}
  \text{BCE}(y, \hat{y}, w) = w_{0} y \log(\hat{y}) + w_{1} (1 - y)
  \log (1 - \hat{y})
,\end{equation}

\noindent where $y$ is the target (either 0 or 1), $\hat{y}$ is the
prediction of the model, and $w$ is the loss weight. This function will
have a very low value when the confidence that the model has in the
class $y$ is high and a high value otherwise. The goal of the training process is to minimise the value of this function on our training dataset.

We define the \textit{mask loss} $\ell_{\text{m}}$ between a target
mask $m \in \{0, 1\}^{H \times W}$ and model output $\hat{m} \in [0,
  1]^{H \times W}$ as follows:

\begin{equation}
  \ell_{\text{m}}(m, \hat{m}, w) = \frac{1}{HW}
  \sum_{
    i = 1
  }^{H}
  \sum_{
    j = 1
  }^{W}
  BCE(m_{ij}, \hat{m}_{ij}, w)
.\end{equation}

The mask loss is effectively a classification loss at the pixel level
of the diagram. In practice we give much more weight ($w_{1}$) to
the perturbation class as this classification problem is very
unbalanced. In the training dataset, diagrams containing perturbations
only have around 2\% of their pixel tagged as perturbation pixels.

We define the \textit{diagram loss} $\ell_{d}$ between the diagram
class $y$ and the model predicted class $\hat{y}$:

\begin{equation}
  \ell_{\text{d}}(y, \hat{y}, w) = BCE(y, \hat{y}, w)
,\end{equation}

\noindent where $w$ is the loss weight.

The global loss of the model $\ell$ is a weighted sum of the mask loss
and the diagram loss:

\begin{equation}
  \ell(y, \hat{y}, m, \hat{m}, w, w_{\text{m}}, w_{\text{d}}) = w_{0}
  \ell_{\text{m}}(m, \hat{m}, w_{\text{m}}) + w_{1}
  \ell_{\text{d}}(y, \hat{y}, w_{\text{d}}).
\end{equation}

\subsection{Training methodology}

To train the model, we used the \textit{Adam optimizer}
\citep{kingma2014adam} with a learning rate of $10^{-3}$ and a
\textit{batch size} of 45. The Adam algorithm is a variant of the
classical stochastic gradient descent that uses the first and second
moments of the gradients. The batch size is the number of samples that
we used to approximate the gradient of the loss function during each optimisation step.

When training a neural network, \textit{data augmentation} is typically
used to help the model acquire invariants (such as rotation and noise
invariance), virtually increase the diversity of the dataset, and
reduce sample memorisation. It consists in randomly modifying the
input of the model according to the augmentation strategy right before
feeding them to the model. In our case, the data augmentation strategy
consists in horizontal flips and random horizontal cyclical rotation
of the diagrams.

In the training loop we used multiple algorithms to make the best use
of our hardware.

We used a trick called \textit{gradient accumulation}
\citep{thomas2018accumulation} to simulate our batch size of 45 as it
did not fit in our GPU memory. This technique entails performing
multiple forward and backward passes on different batches for each
optimiser step.

We also used an algorithm called \textit{data echoing}
\citep{choi2019faster} to greatly speed up the training process. This
technique involves performing multiple optimisation steps for
each data loading by applying the different data augmentation. As
fetching data from the disk is usually very slow, using this technique
allows us to greatly reduce the impact of diagram loading on the
training run-time. In our training runs, each diagram is used three times
for each loading.

In order to improve the model ability to detect very low-magnitude
signals among noise, we also implemented various
\textit{curriculum learning} strategies. Curriculum learning
strategies consist in gradually increasing the difficulty of the
target task along the training process. We experimented with multiple
ways to modify the task:

\begin{itemize}
  \item Progressively inserting samples of decreasing S/N to the training dataset. In our experiments, this approach
    does not improve the model performance when compared to starting
    the process with all of the sample at once.
  \item Adding a Gaussian noise with a mean of zero and slowly increasing
 the    standard deviation to the model input. We call this algorithm the
    \textit{Gaussian virtual curriculum learning}.
  \item Adding a more realistic noise of increasing amplitude to the
    model input. This noise is created from the input diagram by
    applying random cyclical rotations to each of its rows, scaling
    down the values to the desired amplitude and adding the result
    back to the original river diagram. The idea behind this
    methodology is to add a noise component to our sample that closely
    matches the one we could expect from this type of star. By rotating
    the rows by different amounts, we ensure that we are only left
    with noise. We call this algorithm the \textit{star virtual
      curriculum learning}.
\end{itemize}

\subsection{Experimentations}
\label{sec:experiments}
We trained models with and without using data augmentation. The
overfitting phenomenon (in which the model metrics are much better on
its training set than on its validation set) is greatly reduced by the
data augmentation, and therefore  we use it systematically.

The data-echoing algorithm increases the speed of the
training process significantly. As mentioned in the main text, the variety of data used to approximate the gradients of
the loss function is reduced, the quality of the approximation is
therefore also reduced. By having a data augmentation strategy that
greatly modifies a sample, this loss of quality does not significantly
impact the training performance.

The virtual curriculum learning strategies have provided important
metric improvements on previous model and dataset versions. We have
yet to get such gains on our latest models. More hyperparameter tuning
is needed to conclude whether this methodology can yield a boost in
performance or not.

\begin{figure}
  \begin{center}
    \includegraphics[width = 9cm]{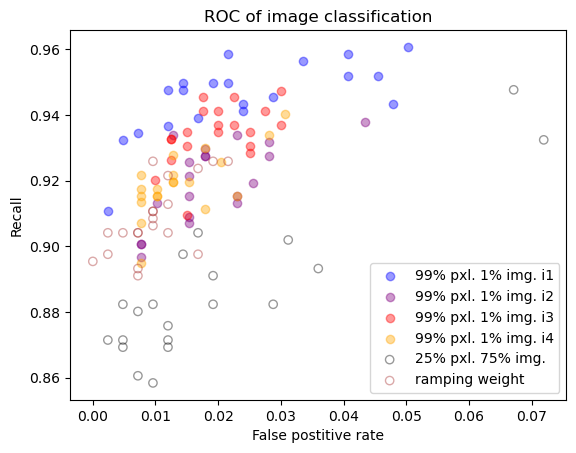}
  \end{center}
  \caption{Receiver operating characteristic of the diagram
    classification task for different global loss weighting
    parameters. For each model, the metrics of the last 20 epochs are
    reported.}
  \label{fig:roc_img}
\end{figure}

\begin{figure}
  \begin{center}
    \includegraphics[width = 9cm]{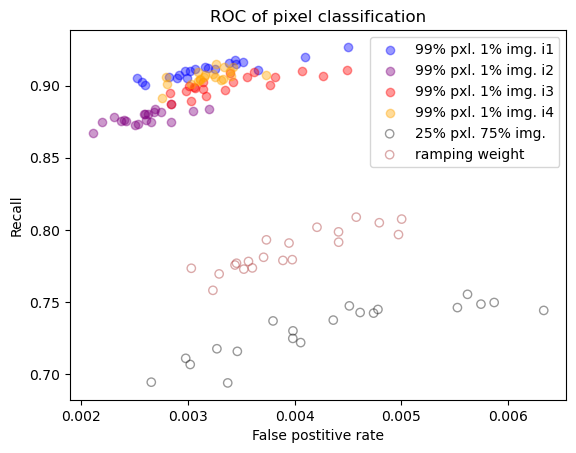}
  \end{center}
  \caption{Receiver operating characteristic of the pixel
    classification task for different global loss weighting
    parameters. For each model, the metrics of the last 20 epochs are
    reported.}
  \label{fig:roc_pixel}
\end{figure}

We have also tried multiple loss weighting strategies. Our experiments suggest that the bigger the weight on the semantic segmentation task the best the overall
performances. Figures~\ref{fig:roc_img} and~\ref{fig:roc_pixel}
illustrate the impact of various global loss weighting parameters. In
these figures, the loss weighting is the following:
\begin{itemize}
\item For the `25\% pxl 75\% img' model, the weight of the mask loss
  $w_{m}$ and the diagram loss $w_{d}$ are respectively 0.25 and 0.75.
\item For the `ramping weight' model, $w_{m}$ starts at a value of
  0.01 and linearly increases to 0.75 from the epoch 10 to 30. At each
  step, the loss of the diagram loss is $w_{d} = 1 - w_{m}$.
\item For all the `99\% pxl 1\% img' models, $w_{m} = 0.99$ and
  $w_{d} = 0.01$.
\end{itemize}

\section{Planetary spin dynamics}  
\label{sec:app_spin}

\begin{figure}
  \begin{center}
     \includegraphics[width = 9cm]{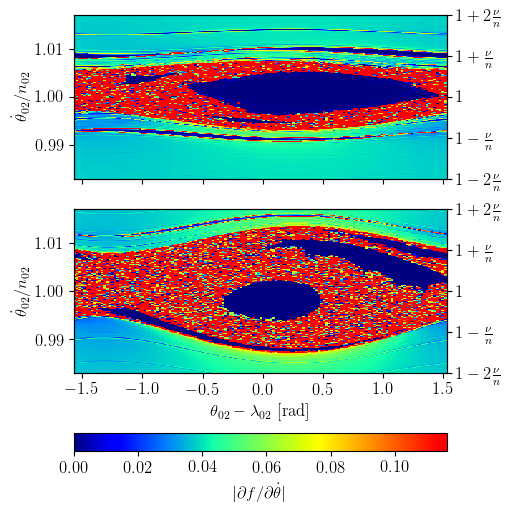}
  \end{center}
  \caption{Spin evolution of Kepler-1705b for different values of the C22 coefficient. \textit{Top:} Residual C22. \textit{Bottom:} Synchronous averaged C22. Each pixel represents an initial condition for the spin of the planet. The x-axis shows the initial position of the ellipsoid with respect to the direction of the star, while the y-axis represent the initial spin rate. The colour of the pixels shows the derivative of the main frequency of the spin of the planet ($f$) with respect to the initial spin rate. This is an indicator of the spin state \citep{LeRoCo2016}: dark blue indicates a libration inside a spin-orbit resonance, cyan/green indicates a circulation of the spin, and red area indicates chaotic rotation.}
  \label{fig:spin_in}
\end{figure}

\begin{figure}
  \begin{center}
     \includegraphics[width = 9cm]{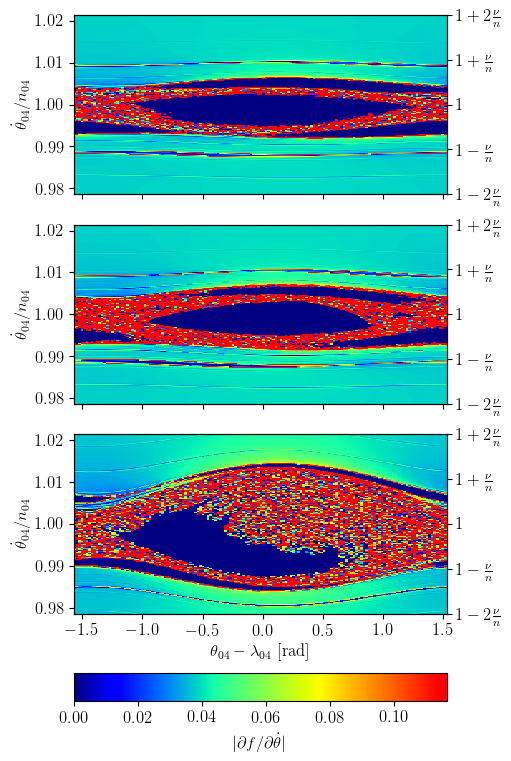}
  \end{center}
  \caption{Spin evolution of Kepler-1705c for different values of the C22 coefficient, see Fig. \ref{fig:spin_in} for explanations. \textit{Top:} Residual C22. \textit{Middle:} Asynchronous averaged C22. \textit{Bottom:} Synchronous averaged C22. }
  \label{fig:spin_out}
\end{figure}

In this Appendix, we study the impact of the planet--planet perturbations on the spin
dynamics of Kepler-1705b and Kepler-1705c, following the method described in \citet{DeCoLeRo2017}.
Denoting the planet's rotation angle with respect to an inertial
line by $\theta$,
the centre of the synchronous resonance is located at $\dot\theta \approx n$,
while subsynchronous and super-synchronous resonances
are located at $\dot\theta \approx n \pm \frac{\nu}{2}$, 
where $n$ is the planet mean-motion and $\nu$ is the perturbation's angular frequency \citep{CoRo2013}.
The frequency $\nu$ is also the dominant frequency observed in the TTVs.
The qualitative evolution of the spin depends on the width of the synchronous and sub or super-synchronous resonances, as well as the distance between them \citep{1979CB}.
The width of the synchronous resonance, $\sigma$,
depends on the planet's asymmetry,
and in particular on the $C_{22}$ Stokes gravity field coefficient.
The width of the sub/super-synchronous resonances is of the order of $\sqrt{\alpha} \sigma$, where $\alpha$ is the TTVs angular amplitude \citep[see][]{DeCoLeRo2017}.

We estimate the $C_{22}$ coefficients of  Kepler-1705b and c following the method
described in \citet{DeCoLeRo2017} (Appendices~C and D).
The planet asymmetry has two contributions, the permanent (or residual) deformation $C_{22,r}$
and the tidally induced deformation $C_{22,t}$.
The permanent deformation can be roughly estimated from a scaling law obtained from
observations of the Solar System planets \citep{yoder_1995_astrometric}
\begin{equation}
    C_{22,r} \sim 10^{-6} \left(\frac{R}{R_{\oplus}}\right)^{5} \left(\frac{M}{M_{\oplus}}\right)^{-5/2}.
\end{equation}
As there are no analogues of  Kepler-1705b or c in the Solar System,
this scaling law only provides a very crude approximation of the permanent deformation
of these planets, which we use in the absence of a more educated guess.
We obtain $C_{22,r}\sim 8\times 10^{-7}$ for  Kepler-1705b
and $C_{22,r}\sim 5\times 10^{-7}$ for  Kepler-1705c.
The tidally induced deformation originates from the adjustment of the planet's mass distribution
to the external gravitational potential.
The deformation of the planet depends on its internal structure and in particular
its relaxation time.
If the deformation were instantaneous, the planet's shape would follow the
variations of the external gravitational potential.
On the contrary, if the relaxation time is much longer than the timescale
of the gravitational potential variations, 
the deformation follows the average external gravitational potential.
Outside of spin-orbit resonances, the tidal deformation averages out,
but if the spin is captured in a resonance, the tidal deformation builds up.
The average deformation depends on the considered spin-orbit resonance \citep{CoRo2013}.
Following \citet{DeCoLeRo2017}, we compute the average $C_{22,t}$
for the synchronous and sub- or super-synchronous resonances,
taking into account the amplitude of forced oscillations.
The total $C_{22}$ is then the sum of the permanent and tidally induced contributions.
For  Kepler-1705b, we obtain $C_{22}\approx 3.5\times 10^{-6}$ in the synchronous resonance,
while in the sub- and super-synchronous resonances, the contribution of the tidally induced deformation is negligible {\bf($C_{22,t} \ll C_{22}\approx C_{22,r} = 8\times 10^{-7}$)}.
For  Kepler-1705c, we obtain $C_{22} \approx 4.4\times 10^{-6}$ in the synchronous resonance
and $C_{22} \approx 8\times 10^{-7}$ in the sub- and super-synchronous resonances.

We plot in Fig.~\ref{fig:spin_in} the phase portraits of the spin of  Kepler-1705b using the estimated $C_{22,r}$ (top) and total $C_{22}$ in the synchronous resonance (bottom). These maps are obtained by n-body integration of the resonant pair and their star as point mass, in addition to the spin of one of the planets represented as an ellipsoid; see section 4.3 of \cite{LeRoCo2016} for more details.
In the case of the permanent deformation alone, we observe a single large stable area corresponding to the synchronous resonance.
This stable area is surrounded by a chaotic region which arises from the overlap of the synchronous
resonance with the two sub- and super-synchronous resonances.
When additionally accounting for the tidally induced deformation in the synchronous resonance,
the chaotic area increases and the stable area shrinks but still exists.
A permanent capture in the synchronous resonance is the most probable scenario
for this planet, while a capture in the non-synchronous resonances does not seem possible.
However, given the size of the chaotic area,
the spin of  Kepler-1705b could undergo a chaotic evolution for a long time before
tidal dissipation finally brings it to permanent capture in the synchronous resonance
\citep[see][]{DeCoLeRo2017}.

The phase portraits of the spin of  Kepler-1705c are shown in Fig.~\ref{fig:spin_out}.
For this planet, we plot three phase portraits, 
corresponding to the permanent deformation alone (top),
the total deformation in the sub- or super-synchronous resonances (middle),
and the total deformation in the synchronous resonance (bottom).
Considering only the permanent deformation, 
we observe three stable areas corresponding to the three resonances (synchronous, sub-, and super-synchronous).
The three resonances slightly overlap, which generates a chaotic area around the separatrices
of these resonances.
The three stable areas survive when taking into account the tidally induced deformation in the sub- and super-synchronous resonances.
This suggests that a permanent capture in these non-synchronous resonances is possible for  Kepler-1705c.
Finally, when taking into account the tidally induced deformation in the synchronous resonance,
the chaotic area extends and only the stable island corresponding to the synchronous resonance remains.
Therefore, for this planet, permanent capture in any of the three resonances is possible.
As for  Kepler-1705b, the spin of  Kepler-1705c could also undergo a chaotic evolution for a long time before
being permanently captured in one of these resonances.
The probability of capture in each resonance should be roughly proportional to its area in Fig.~\ref{fig:spin_out} (top).
The sum of the areas of the two non-synchronous resonances is of the same order as the area of the synchronous resonance, which suggests that the spin of  Kepler-1705c has a significant probability of being currently in a non-synchronous state.
A more precise estimate of this probability would require a detailed study
with long-term simulations including tidal dissipation,
as was done in \citet{DeCoLeRo2017},
which is beyond the scope of this article.

\end{document}